\documentclass[a4paper,twoside,12pt]{report}

\usepackage{amsmath,exscale,amsthm}
\usepackage{latexsym,fancyhdr}
\usepackage{graphicx,psfrag}
\usepackage[small,bf,hang]{caption}
\usepackage{cite}
\begin{document}
%
%
\thispagestyle{empty}

\newcommand{\thesisTitle}{Analysis of a Shell System in General Relativity}

  \vspace{4cm}

\noindent
  \parbox{\textwidth}{\huge\bf\thesisTitle}    
  \vspace{3cm}

  \noindent
  Inauguraldissertation \\
  der Philosophisch-naturwissenschaftlichen Fakult\"at \\
  der Universit\"at Bern

  \vspace{1.5cm}

  \noindent
  vorgelegt von

  \vspace{4mm}

  \noindent  {\large\bf Fabio Scardigli}

  \vspace{3mm}

  \noindent
  von Milano, Italia

  \vspace{1.5cm}

  \noindent
  Leiter der Arbeit: \parbox[t]{7cm}{Prof. Dr. Petr H\'{a}j\'{\i}\v{c}ek \\
    Institut f\"ur theoretische Physik \\
    Universit\"at Bern}
  
  \vspace{6cm}

  \thispagestyle{empty}

  \noindent
  \parbox{\textwidth}{\huge\bf\thesisTitle}

  \vspace{2cm}

  \noindent
  Inauguraldissertation \\
  der Philosophisch-naturwissenschaftlichen Fakult\"at \\
  der Universit\"at Bern

  \vspace{1.0cm}

  \noindent
  vorgelegt von

  \vspace{4mm}

  \noindent
  {\large\bf Fabio Scardigli}

  \vspace{3mm}

  \noindent
  von Milano, Italia

  \vspace{1.5cm}

  \noindent
  Leiter der Arbeit: \parbox[t]{7cm}{Prof. Dr. Petr H\'{a}j\'{\i}\v{c}ek \\
    Institut f\"ur theoretische Physik \\
    Universit\"at Bern}

  \vspace{1.5cm}

  \noindent
  Von der Philosophisch-naturwissenschaftlichen Fakult\"at angenommen.

  \vspace{1.5cm}

  \noindent
  Bern, den 20. Dezember 2001 
  \hfill
  \parbox[b]{5cm}{Der Dekan:
    \\ \vspace{2.5cm} \\
    Prof. Dr. P. Bochsler}


%
\setcounter{page}{1}
\pagenumbering{roman}
\tableofcontents
\setcounter{page}{1}
\pagenumbering{arabic}
%
%
\chapter{Introduction}
One of the problems that seems to be crucial for the development of theoretical 
physics is, surely, the one of gravitational collapse. This problem has become, 
during the decades, of central importance not only for the comprehension of
high energy astrophysical phenomena, but also as a key point for the correct 
formulation itself of the theory of gravity. The gravitational collapse, with
the appearence of a singularity, leads to serious problems in classical general 
relativity ~\cite{Hawking}. The structure of the resulting singularity contradicts 
the axioms 
of the  theory such as the equivalence principle. Thus, the proof of the 
singularity theorems was interpreted by many researchers as the unavoidable 
necessity of a quantum theory of gravity. It is in fact a general expectation 
that a quantum theory of gravity can lead to a singularity-free geometry. \\
This goal, the Holy Graal of Theoretical Physics, has been pursued during the 
years in at least two ways. On one side, one can try to construct a full theory
of gravity and then derive from it the correct behaviour of collapsing matter 
(examples of this line of research are the superstring theory or the loop 
quantum gravity). On the other side, one can attemp to build models (hopefully, 
simple models!) which can be exactly quantized and which therefore serve as a 
guide towards the full, unknown, theory. This aspect of the situation reminds
the first models on atomic structure which led to very good predictions and
insights without the knowledge of full quantum electrodynamics. Thus, impatient 
people try to capture the core of the problem by working with semplified models.\\
And the simplest models of the collapse are the spherically symmetric thin shells
coupled to their own gravitational field ~\cite{PH1}, ~\cite{PH2}, ~\cite{PH3}. 
The classical dynamics of objects with discontinuities in matter density is well 
understood ~\cite{Israel}, ~\cite{MTW}.
Spherically symmetric thin shells 
are popular models used extensively in a number of phenomena: classical 
gravitational collapse ~\cite{Cruz}, entropy of black holes ~\cite{York}, 
quantum theory of black holes ~\cite{Casadio}, 
quantum theory of gravitational collapse ~\cite{PH2}. In the last problem, for
example, surprising results have been obtained in ~\cite{PH4}, ~\cite{PH5},
~\cite{PH6}.\\
A generalization of these models, which appears quite natural, is to consider
thin shells without any symmetry, in particular with no spherical symmetry.
The paper ~\cite{HK} (in the following, HK paper), whose the present 
work aims
to represent a development and (hopefully!) a clarification of some aspects, 
contains a 
reformulation of the dynamics of gravitation together with ideal discontinuos
fluid, in hamiltonian form. Precisely, a barotropic ideal fluid with
$\delta$-function discontinuities (i.e. fluid thin shell) is coupled to Einstein
gravity in hamiltonian formalism ~\cite{JK1}, ~\cite{JK2}, ~\cite{JK3}. 
It is shown, in ~\cite{HK}, 
that the constraints 
and the canonical equations resulting from this hamiltonian approach are 
equivalent to the system of Einstein equations plus dynamical equations of 
ideal  fluid, plus Israel equations (in case of thin shells).\\
An interesting problem arises in the study of this model: to show that the 
hamiltonian formalism defines a (regular) constrained system ~\cite{Dirac}. This means,
in particular, that the hamiltonian and the constraints must be  differentiable 
functions on the phase space so that their Poisson bracket are well defined objects.
This is a key starting point for the quantization of the model.
The difficulty pointed out at the end of the HK paper is that some constraints
at the shell are not differentiable functions on the phase space.
The best way to takle this problem is (following Teitelboim and Henneaux 
~\cite{Teitel})
to solve the singular constraints and to substitute the solution back into the 
action (in our case, back into the hamiltonian ~\cite{JK1}). So that the resulting 
variational principle leads to equivalent dynamics without singular constraints.\\
The aims of the present thesis are to eliminate these singular constraints, using
the reduction procedure suggested by Teitelboim and Henneaux, and to show 
the full differentiability of the reduced system.\\
These goals have been fully achieved in the present work.\\
The singular constraints $\gamma$-C, $\alpha$-C, $\lambda$-C have been solved 
in sequence (see the related chapters). 
At every step there is a progressive reduction
of the number of variables.  
After every solution the corresponding new variables, 
new hamiltonian, new phase space, new symplectic form, new equations of 
motion and constraints are clearly stated. It is particularly interesting the 
$\gamma$-C, which contains a great amount of geometrical information.
In the end we are left with the equations of motion only 
(plus canonical volume and surface constraints), and the singular 
constraints $\gamma$-C, $\alpha$-C, $\lambda$-C have been completely eliminated.
The equations of motion are equivalent to the starting ones, although they are 
of course expressed in the variables remaining after the solution of the singular
constraints. Finally, chapter 5 is devoted to show the differentiability of 
the resulting hamiltonian and of the final canonical constraints 
on the reduced phase space.     
%
%
%
\chapter{The $\gamma$ Constraint}
The present section describes the geometrical environment where our 
considerations take place and the identification of the singular constraints 
which plague our system. Then, we show how a particular choice of coordinates 
enable us to "eliminate" the gamma-constraint ($\gamma$-C in the following).
\section{Geometrical Environment}
We remind briefly the geometrical set up in which we operate; 
it is widely described in ~\cite{HK} (HK paper). 
We call $\Sigma$ the timelike surface swept 
by the shell during its motion ($dim\Sigma=3$) in the spacetime $M$ ($dimM=4$). 
$\Sigma$ cuts $M$ in two volumes, so that $M=V^{-} \cup V^{+}$. A foliation of 
these volumes is introduced via an arbitrary family of spacelike 
Cauchy surfaces $S^{+}$, $S^{-}$. 
We admit also, in general, that such surfaces can have  
a cusp at the intersection $S \cap \Sigma$. 
$S \cap \Sigma$ is the shell ($dim S \cap \Sigma =2$).
We introduce also some other geometrical objects, in 
order to better characterize our system. Namely, the normals:

\begin{tabbing}
$n^{\mu}$ is the normal to $S$ in $M$ \\
$m^{\mu}$ is the normal to $S\cap\Sigma$ in $S$ ($\textbf{m}\perp\textbf{n})$ \\
$\tilde{n}^{\mu}$ is the normal to $S\cap\Sigma$ in $\Sigma$ \\
$\tilde{m}^{\mu}$ is the normal to $\Sigma$ in $M$ ($\tilde{\textbf{m}}\perp
\tilde{\textbf{n}})$.
\end{tabbing}
And the angle $\alpha$ such that $\sinh\alpha := -\textbf{n}\cdot\tilde{\textbf{m}} = 
-g_{\mu\nu}n^{\mu}{\tilde{m}}^{\nu}$.\\ 
The indexes take the values: $\mu,\nu=0,1,2,3$; $\alpha,\beta=0,1,2$; $k,l=1,2,3$;
$A,B=1,2$.  
Quantities as the normal $n^{\mu}$ 
to $S$ in $M$, the angle $\alpha$, etc., can have a step 
discontinuity at $S \cap \Sigma$. The surface $\Sigma$ is the world volume 
swept by the fluid shell in the spacetime $M$. 
The description of our system, "Gravity $\oplus$ Fluid Shell", starts from 
the action written in surface form (see formula 94 of HK):

\begin{equation}
\begin{align}
S & =  \frac{1}{16 \pi G} \int_{V^{-}}d^{4}x\sqrt{|g|}R + \frac{1}{16 \pi G} 
\int_{V^{+}}d^{4}x\sqrt{|g|}R - \frac{1}{8 \pi G}\int_{\Sigma}d^{3}
\xi\sqrt{|\gamma|}[L] \nonumber \\ \nonumber \\ 
& -  \int_{\Sigma}d^{3}\xi\sqrt{|\gamma|}n_{s}e_{s}(n_{s}).
\label{S1}
\end{align} 
\end{equation}
The first two terms on the RHS are the volume part of the gravitational action, 
the third term is the shell part of the gravitational action, and the last one is 
the matter action. The fields in the action S are the gravity field 
$g_{\mu\nu}(x)$ in $V^{\pm}$ (=metric of the spacetime $M$); the gravity field 
$\gamma_{\alpha\beta}(y)$ in $\Sigma$ (=metric on $\Sigma$); the curvature scalar 
$R$; $L$ is the second fundamental form of $\Sigma$ in $M$, $L=\gamma^{\alpha\beta}
L_{\alpha\beta}$, where 
$L_{\alpha\beta}=\tilde{m}_{\mu;\nu}e^{\mu}_{\alpha}e^{\nu}_{\beta}$ 
(square brackets [...] mean jump at $\Sigma$); 
$n_{s}$ is the surface mole density; $e(n_{s})$ is the energy per mole.
Of course $n_{s}=n_{s}(z^{A})$, where $z^{A}$ are the matter fields on $\Sigma$.    
All the integrands in this action are smooth functions (not distributions) and 
S can be used with completely general 
coordinates, namely, arbitary smooth coordinates $x_{\pm}^{\mu}$
within $V^{\pm}$ and arbitrary smooth coordinates $\xi^{\alpha}$ within 
$\Sigma$. 
\section{Continuity Relation}
The metric is required to satisfy the 
"continuity relation" (called $\gamma$-C):
\begin{equation}     
(g_{\mu\nu}e^{\mu}_{\alpha}e^{\nu}_{\beta})^{+} = 
(g_{\mu\nu}e^{\mu}_{\alpha}e^{\nu}_{\beta})^{-} = \gamma_{\alpha\beta}(\xi) 
\label{gammaC}
\end{equation}  
where $e^{\mu}_{\alpha}=\partial x^{\mu}/\partial \xi^{\alpha}$ and
the symbols $\pm$ denote here the limits from the volumes   
$V^{\pm}$ towards $\Sigma$. The role of the continuity relation 
(\ref{gammaC}) is to define the configuration space of our system. The relation 
(\ref{gammaC}) is a \textit{constraint} in the sense that the coordinates  
$x_{\pm}^{\mu}$ within $V^{\pm}$ and $\xi^{\alpha}$ 
within $\Sigma$ can be chosen arbitrarily as well as the metric $g_{\mu\nu}(x)$ 
in $V^{\pm}$ and $\gamma_{\alpha\beta}(y)$ on $\Sigma$, but the 
arbitrariety is not complete, because the relation (\ref{gammaC}) must hold. 
This means also that the sector $(012) \times (012)$ of the metric 
$g_{\mu\nu}$ must be continuos across $\Sigma$. 

\section{Introduction of Adapted Coordinates}
\label{secAC}
Following the HK paper we introduce now, in the preceding geometrical 
situation, a system of coordinates $x^{\mu}$ adapted to the geometry. 
The Cauchy surfaces $S_{t}$ are given by $x^{0}=t$, and $\Sigma$ by $x^{3}=0$.
Besides we put: $x^{k}|_{S_{t}} =: y^{k}$, ($k=1,2,3$),  $x^{\alpha}|_{\Sigma} =: 
\xi^{\alpha}$, ($\alpha=0,1,2$),
$x^{A}|_{S\cap\Sigma} =: \eta^{A}$, ($A=1,2$).
We note that this system is "comoving" in the sense that the shell is always 
described, at any time, by the equation $x^{3}=0$. The coordinates are adapted 
to the geometrical situation, and a cusp is allowed in the Cauchy surfaces at
$S\cap\Sigma$. The whole four metric $g_{\mu\nu}(x)$ can also be 
\textit{not continuos} across $\Sigma$, in respect to the introduced 
coordinates. 
That is to say, the relation (\ref{gammaC}) requires the continuity of a part 
of the metric $g_{\mu\nu}$, but not of the whole $g_{\mu\nu}$.
We introduce also the standard \textit{lapse-shift} decomposition.
If $\textbf{e}_{0}=\frac{\partial}{\partial t}$ represents the "deformation vector", 
i.e. the vector field 
tangent to the coordinate line $x^{0}$; $\textbf{e}_{k}$ ($k=1,2,3$) are the 
tangent vectors to the lines $x^{1}$, $x^{2}$, $x^{3}$; $\textbf{e}_{L}$ 
($L=1,2$) are the tangent vectors to the lines $\eta^{1}$, $\eta^{2}$ 
(we are in adapted coordinates, therefore we have $e^{k}_{L} = 
\frac{\partial y^{k}}{\partial \eta^{L}} = \frac{\partial x^{k}}
{\partial x^{L}}$ ); then we can write the decompositions:

\begin{tabbing}
$\textbf{e}_{0}=N\textbf{n} + N^{k}\textbf{e}_{k}$ \\ \\
$\textbf{e}_{0}=\nu\tilde{\textbf{n}} + \nu^{L}\textbf{e}_{L}$.
\end{tabbing}
Note that if a cusp is present on S at $S\cap\Sigma$, then the jumps 
at $S\cap\Sigma$ are $[\textbf{n}]\neq 0$, $[\alpha]\neq 0$, 
$[\textbf{m}]\neq 0$ (while, of course, $[\tilde{\textbf{n}}]=
[\tilde{\textbf{m}}]=0$). This means that, in this case the relation 
$\textbf{e}_{0}=N\textbf{n} + N^{k}\textbf{e}_{k}$ has two different forms 
in $V^{-}$ and in $V^{+}$. Namely:

\begin{tabbing}
from the $V^{-}$ side: $\textbf{e}_{0}=N^{-}\textbf{n}^{-} + 
N^{k}_{-}\textbf{e}_{k}^{-}$ \\ \\
from the $V^{+}$ side: $\textbf{e}_{0}=N^{+}\textbf{n}^{+} + 
N^{k}_{+}\textbf{e}_{k}^{+}$
\end{tabbing}
We have: $\textbf{n}^{+}\neq\textbf{n}^{-}$, $\textbf{e}^{+}_{3}\neq
\textbf{e}^{-}_{3}$, $\textbf{e}^{+}_{2}=\textbf{e}^{-}_{2}$,
$\textbf{e}^{+}_{1}=\textbf{e}^{-}_{1}$.
We remind also two important relations among the normals and the angle 
$\alpha$ (see HK paper for a proof of such relations):

\begin{tabbing}
$\tilde{m}^{\mu}= n^{\mu}_{\pm}\sinh\alpha_{\pm}+ 
m^{\mu}_{\pm}\cosh\alpha_{\pm}$ \\ \\
$\tilde{n}^{\mu}= n^{\mu}_{\pm}\cosh\alpha_{\pm}+ 
m^{\mu}_{\pm}\sinh\alpha_{\pm}$
\end{tabbing} 
We are now able to obtain two important relations among $N$, $N^{k}$, 
$\nu$, $\alpha$. We have to distinguish, of course, the version in $V^{-}$
from that in $V^{+}$. In order to do not make cumbersome the notation, we 
drop the indexes $\pm$ during the calculations and we restore them only at 
the end. From the decompositions of $\textbf{e}_{0}$ we can write:

\begin{equation}     
N\textbf{n} + N^{k}\textbf{e}_{k}=\nu\tilde{\textbf{n}} + \nu^{L}\textbf{e}_{L}
\end{equation}
Multiplying the last equation first by $\textbf{m}$ and then by $\textbf{n}$
we get:
\begin{equation}
\begin{align}     
& N\textbf{n}\cdot\textbf{m} + N^{k}\textbf{e}_{k}\cdot\textbf{m}=
\nu\tilde{\textbf{n}}\cdot\textbf{m} + \nu^{L}\textbf{e}_{L}\cdot\textbf{m}
\quad\Rightarrow \nonumber \\ 
& N^{\perp}=\nu(\textbf{n}\cdot\textbf{m}\cosh\alpha+
\textbf{m}\cdot\textbf{m}\sinh\alpha) = \nu\sinh\alpha
\end{align}
\end{equation}

\begin{equation}
\begin{align}     
& N\textbf{n}\cdot\textbf{n} + N^{k}\textbf{e}_{k}\cdot\textbf{n}=
\nu\tilde{\textbf{n}}\cdot\textbf{n} + \nu^{L}\textbf{e}_{L}\cdot\textbf{n}
\quad\Rightarrow \nonumber \\ 
-& N=\nu(\textbf{n}\cdot\textbf{n}\cosh\alpha+
\textbf{m}\cdot\textbf{n}\sinh\alpha) = -\nu\cosh\alpha
\quad\Rightarrow \nonumber \\
& N=\nu\cosh\alpha
\end{align}
\end{equation}

We have used the properties:\\ \\
\( \begin{array}{ll}
\textbf{n}\cdot\textbf{m}=0 & \quad \quad\quad \textbf{e}_{L}\cdot\textbf{n}=0  \\
\textbf{e}_{L}\cdot\textbf{m}=0 & \quad\quad\quad\textbf{e}_{k}\cdot\textbf{n}=0  \\
N^{k}\textbf{e}_{k}\cdot\textbf{m}=N_{k}m^{k}=N^{\perp} & \quad\quad\quad 
\textbf{n}\cdot\textbf{n}=-1  \\
\textbf{m}\cdot\textbf{m}=1
\end{array} \) \\ \\
Restoring the indexes $\pm$, we can write:
\begin{tabbing}
$N^{\perp}_{\pm}=\nu\sinh\alpha_{\pm}$ \\
$N_{\pm}=\nu\cosh\alpha_{\pm}$
\end{tabbing}
We note that we have obtained such relations from the pure geometry 
(the choice of adapted coordinates)
and not as consequences of the dynamics (so far not yet introduced).

\section{Introduction of the Hamiltonian $\mathcal{H}_{118}$} 
\label{secH118}

In this section we give the form of the hamiltonian producing the 
evolution of our system. We mantain the use of the index ($118$) as 
in the HK paper. This hamiltonian can be derived from the action 
(\ref{S1}) (see HK paper for explicit passages). 
Using the adapted coordinates before defined,
the hamiltonian is written as (see HK paper, formula 108)

\begin{equation}
\begin{align}
\mathcal{H} & = \frac{1}{8 \pi G} \int_{S^{-}\cup S^{+}}d^{3}y G^{0}_{0} +
 \frac{1}{8 \pi G} \int_{S\cap\Sigma}d^{2}\eta[Q_{0}^{0}]-\int_{S\cap\Sigma}
 d^{2}\eta T^{0}_{s\;0}\nonumber \\ \nonumber \\ 
 & +\frac{1}{8 \pi G} \int_{S\cap\Sigma^{+}}d^{2}\eta\sqrt{|\gamma|}L^{0}_{0}
\label{H108}
\end{align}
\end{equation}
The well known decompositions hold
\begin{equation}
\begin{align}
G^{0}_{0} & =\frac{1}{2}(NC+N^{k}C_{k}) \nonumber \\
[Q_{0}^{0}] & = -\frac{\sqrt{\lambda}}{\sqrt{|\gamma|}}(\nu[Q^{\perp\perp}]+
\nu^{K}[Q^{\perp}_{K}]) \nonumber \\
T^{0}_{s\;0} & = -\frac{\sqrt{\lambda}}{\sqrt{|\gamma|}}(\nu T^{\perp\perp}_{s}+
\nu^{K}T^{\perp}_{s\;K})
\end{align}
\end{equation}
where
\[ \left\{ \begin{array}{ll}
C=\frac{2\pi^{kl}\pi_{kl}-\pi^{2}}{2\sqrt{q}}-\sqrt{q}R^{(3)} \\ \\
C_{k}=-2\pi^{l}_{k|l}
\end{array}
\right.  \quad \left\{ \begin{array}{ll} 
\frac{Q^{\perp\perp}}{\sqrt{|\gamma|}}=
\tilde{\pi}^{\perp\perp}\sinh\alpha-l\cosh\alpha \\ \\

\frac{Q^{\perp}_{K}}{\sqrt{|\gamma|}}= 
\tilde{\pi}_{K}^{\perp}+\alpha_{,K}
\end{array}
\right.    \quad \left\{ \begin{array}{ll}
T^{\perp\perp}_{s}=T^{\alpha\beta}_{s}\tilde{n}_{\alpha}\tilde{n}_{\beta}  \\ \\
T^{\perp}_{s\;K}=T^{\alpha\beta}_{s}\tilde{n}_{\alpha}e_{\beta K}
\end{array}
\right.    \] \\ \\
and $T^{\alpha\beta}_{s}$ is the surface stress-energy tensor of the shell.\\
Therefore, the hamiltonian becomes
\begin{equation}
\begin{align}
\mathcal{H}_{118} & = \frac{1}{16 \pi G} \int_{S^{-}\cup S^{+}}d^{3}y 
\left\{N\left(\frac{2\pi^{kl}\pi_{kl}-\pi^{2}}{2\sqrt{q}}-\sqrt{q}R^{(3)}\right)+
N^{k}(-2\pi^{l}_{k|l})\right\} \nonumber \\ 
& -\frac{1}{8 \pi G} \int_{S\cap\Sigma}d^{2}\eta
\sqrt{\lambda}(\nu[\tilde{\pi}^{\perp\perp}\sinh\alpha-l\cosh\alpha]+\nu^{K}
[\tilde{\pi}_{K}^{\perp}+\alpha_{,K}]) \nonumber \\
& -\int_{S\cap\Sigma}
 d^{2}\eta T^{0}_{s\;0}+\frac{1}{8 \pi G} \int_{S\cap\Sigma^{+}}d^{2}\eta
 \sqrt{|\gamma|}L^{0}_{0}
\label{H118}
\end{align}
\end{equation}
The integrals over $S\cap\Sigma^{+}$ vanish when $\Sigma^{+}\rightarrow\infty$
and therefore they don't give any contribute to the final equations. Hence they 
will not be re-written anymore. We remind the meaning of several symbols:\\
$N$, $N^{k}$, $\nu$, $\nu^{L}$ are lagrangian multipliers with the known 
geometrical interpretations;\\
$q_{kl}$ is the metric on $S^{\pm}$, projection of $g_{\mu\nu}$ on $S$;\\
$\lambda_{KL}$ is the metric on $S\cap\Sigma$, projection of 
$\gamma_{\alpha\beta}$ on $S\cap\Sigma$;\\
$l=\lambda^{KL}l_{KL}=\lambda^{KL}(m_{k|r}e^{k}_{K}e^{r}_{L})$ is the second 
fundamental form of $S \cap \Sigma$ in $S$;\\
$\tilde{\pi}^{\perp\perp}=(\pi^{kl}m_{k}m_{l})/\sqrt{q}$;\\ 
$\tilde{\pi}^{\perp}_{K}=(\pi^{kl}m_{k}e_{l\,K})/\sqrt{q}$. \\ \\
\underline{Phase space}: We now look at the phase space described by the 
hamiltonian $\mathcal{H}_{118}$. To identify correctly the phase 
space variables, we have to remind the expression of the hamiltonian 
$\mathcal{H}$ as a function of the lagrangian $\mathcal{L}$,
in adapted coordinates (see HK paper, formula 106):
\begin{equation}
\begin{align}
\mathcal{H}= & -\frac{1}{16 \pi G}\int_{S}d^{3}yq_{kl}\dot{\pi}^{kl}-
\frac{1}{8 \pi G}\int_{S\cap\Sigma}d^{2}\eta\sqrt{\lambda}[\dot{\alpha}]\nonumber \\
& +\int_{S\cap\Sigma}d^{2}\eta p_{A}\dot{z}^{A}+
\frac{1}{8 \pi G}\int_{S\cap\Sigma^{+}}d^{2}\eta\sqrt{\lambda}\dot{\alpha}
-\mathcal{L}
\end{align}
\end{equation}
Here $\mathcal{L}$ is the lagrangian. The integral over $S\cap\Sigma^{+}$ vanishes 
when $\Sigma^{+}\rightarrow\infty$,
therefore is not relevant for the correct identification of the conjugate momenta.
From here is clear that the canonical variables sweeping the phase space (and 
the corresponding conjugate momenta) are, in adapted coordinates,\\

\( \begin{array}{cll}
\hspace*{3cm}-\frac{1}{16 \pi G}\;q_{kl} & \longleftrightarrow & \pi^{kl} \\ \\
\hspace*{3cm}-\frac{1}{8 \pi G}\sqrt{\lambda} & \longleftrightarrow & [\alpha] \\ \\
\hspace*{3cm}z^{A} & \longleftrightarrow & p_{A}
\end{array} \) \\ \\
This can be seen also by inspecting the symplectic form
\begin{multline}
\delta\mathcal{H}^{SYM}_{107} = -\frac{1}{16 \pi G}\int_{S}d^{3}y(\dot{\pi}^{kl}
\delta q_{kl}-\dot{q}_{kl}\delta\pi^{kl}) \\
+\frac{1}{16 \pi G}\int_{S\cap\Sigma}d^{2}
\eta\sqrt{\lambda}\left(\frac{\dot{\lambda}}{\lambda}\delta[\alpha]-\dot{[\alpha]}
\frac{\delta\lambda}{\lambda}\right)
-\int_{S\cap\Sigma}d^{2}\eta(\dot{p}_{A}\delta
z^{A}-\dot{z}^{A}\delta p_{A})
\end{multline}
\underline{Equations of motion}: Following a well know general method (see 
HK paper and Kijowsky paper ~\cite{JK2}), 
we vary now the hamiltonian $\mathcal{H}_{118}$
in respect to \textit{all} the variables and we compare this variation with 
the symplectic form $\delta\mathcal{H}^{SYM}_{107}$: we shall obtain, in this
way, equations of motion and constraints. As we will see, the variation will 
be done by thinking all the variables as independent one from the other. $q_{kl}$, 
$\lambda_{AB}$, $[\alpha]$, $N$, $N^{\perp}$, $\nu$, $\nu^{K}$, $\alpha$, are 
all thought as reciprocally independent. We will use the hamiltonian 
$\mathcal{H}_{118}$, written in adapted coordinates, that is with the just 
introduced variables. But we will not take into account, in making the 
variation, the geometrical relations obtained in section \ref{secAC}. 
The variation
of $\mathcal{H}_{118}$, computed by treating the variables as reciprocally 
independent, compared with the symplectic form $\delta\mathcal{H}^{SYM}_{107}$,
will return to us, in the form of constraints, the relations obtained from the geometry in 
section \ref{secAC}.\\ 
As regard the $\gamma$-C, we will see that the adapted coordinates permit 
to build a metric
$g_{\mu\nu}$, all over M, starting from metrics $\gamma_{\alpha\beta}$
assigned on the $x^{3}=const.$ surfaces, in such a way that $g_{\mu\nu}$
satifies the constraint equation (\ref{gammaC}).\\
Now we have a look to the variation $\delta\mathcal{H}_{118}$

\begin{equation}
\begin{align}
\delta\mathcal{H}_{118} & =\frac{1}{16 \pi G}\int_{S}d^{3}y(C_{k}\delta N^{k}+
C\delta N + a^{kl}\delta q_{kl} + b_{kl}\delta\pi^{kl}) \nonumber \\
& +\frac{1}{16 \pi G}
\int_{S\cap\Sigma}d^{2}\eta[2REST-\sqrt{\lambda}B^{k}m_{k}] \nonumber \\
& -\frac{1}{8 \pi G}\int_{S\cap\Sigma}d^{2}\eta\frac{\sqrt{\lambda}}
{\sqrt{|\gamma|}}\left\{\delta\nu\left([Q^{\perp\perp}]-8\pi G T^{\perp\perp}_{s}\right)+
\delta\nu^{K}\left([Q^{\perp}_{K}]-8 \pi GT^{\perp}_{s\;K}\right)\right\} \nonumber \\
& - \int_{S\cap\Sigma}d^{2}\eta\left\{\frac{1}{2}T^{KL}_{s}\delta\lambda_{KL}\right\} 
\nonumber \\
& - \int_{S\cap\Sigma}d^{2}\eta\left\{\left(\frac{\partial T^{0}_{s\;0}}
{\partial z^{A}}-\frac{\partial}{\partial\eta^{M}}\frac{\partial T^{0}_{s\;0}}
{\partial z^{A}_{M}} \right)\delta z^{A}+ \frac{\partial T^{0}_{s\;0}}{\partial 
p_{A}}\delta p_{A}\right\}
\label{varH118}
\end{align}
\end{equation}  \\ \\  
From the comparison $\delta\mathcal{H}_{118}=\delta\mathcal{H}^{SYM}_{107}$
we obtain the equations:

\[ \left\{ \begin{array}{ll}
C_{k}=0 \\ \\
C=0
\end{array}
\right. \quad \left\{ \begin{array}{ll}
\dot{q}_{kl}=b_{kl} \\ \\
\dot{\pi}^{kl}=-a^{kl}
\end{array}
\right.  \quad \left\{ \begin{array}{ll} 
[Q^{\perp\perp}]=8 \pi G T^{\perp\perp}_{s} \\ \\

[Q^{\perp}_{K}]= 8 \pi G T^{\perp}_{s\;K}
\end{array}
\right.    \quad \left\{ \begin{array}{ll}
\dot{p}_{A}=\frac{\partial T^{0}_{s\;0}}
{\partial z^{A}}-\frac{\partial}{\partial\eta^{M}}\frac{\partial T^{0}_{s\;0}}
{\partial z^{A}_{M}} \\ \\
-\dot{z}^{A}=\frac{\partial T^{0}_{s\;0}}{\partial p_{A}}
\end{array}
\right.    \] \\ \\
here the first two couples of equations are the well known Einstein 
equations in canonical form; the third couple of equations represents 
the first three Israel equations;
and the last ones are the dynamical equations for the matter composing the shell
(see HK paper, and also \cite{MTW}). 
In particular, from the comparison 
\begin{equation}
\begin{align}
 & \frac{1}{16 \pi G}
\int_{S\cap\Sigma}d^{2}\eta[2REST-\sqrt{\lambda}B^{k}m_{k}]
-\frac{1}{2} \int_{S\cap\Sigma}d^{2}\eta T^{KL}_{s}\delta\lambda_{KL} = \nonumber \\
  & \frac{1}{16 \pi G}\int_{S\cap\Sigma}d^{2}
\eta\sqrt{\lambda}\lambda^{KL}(-\dot{[\alpha]}\delta\lambda_{KL}
+\dot{\lambda}_{KL}\delta[\alpha])
\end{align}
\end{equation}
which explicitly reads
\begin{equation}
\begin{align}
 & \int_{S\cap\Sigma}d^{2}\eta[-2\sqrt{\lambda}\tilde{\pi}^{\perp\perp}
(\nu\sinh\alpha-N^{\perp})m^{k}m^{l}\delta q_{kl} - \sqrt{\lambda}
\{\tilde{\pi}^{\perp\perp}\lambda^{KL}(\nu\sinh\alpha-N^{\perp}) \nonumber \\
& +\tilde{\pi}^{KL}
N^{\perp}+l^{KL}N-l\lambda^{KL}\nu\cosh\alpha+\nu^{M}\lambda^{KL}\alpha_{,M}
-N_{,k}m^{k}\lambda^{KL}\}\delta\lambda_{KL} \nonumber \\
& +2\sqrt{\lambda}(\nu\cosh\alpha-N)\delta l- 
2\sqrt{\lambda}(\nu\sinh\alpha-N^{\perp})m_{k}m_{l}\delta\tilde{\pi}^{kl} \nonumber \\
& +2\sqrt{\lambda}(-\nu\tilde{\pi}^{\perp\perp}\cosh\alpha+l\nu\sinh\alpha
+\nu^{K}_{||K})\delta\alpha]-\int_{S\cap\Sigma}d^{2}\eta(8 \pi G T^{KL}_{s}
\delta\lambda_{KL})= \nonumber \\
& \int_{S\cap\Sigma}d^{2}
\eta\sqrt{\lambda}\lambda^{KL}(-\dot{[\alpha]}\delta\lambda_{KL}
+\dot{\lambda}_{KL}\delta[\alpha])
\label{1.12}
\end{align}
\end{equation}
we get the equations (reminding that $[A \cdot B]=A^{\pm}[B]+[A]B^{\mp}$) 

\[ \left\{ \begin{array}{ll}
\nu\sinh\alpha_{\pm}-N^{\perp}_{\pm}=0 \\ \\
\nu\cosh\alpha_{\pm}-N_{\pm}=0
\end{array}
\right. \]
We see here that we re-obtain in the form of constraints two equations that 
we obtained in section \ref{secAC} as pure geometric properties of our system.\\
Moreover we get the equation 130HK
\begin{equation}
\begin{align}
& -\sqrt{\lambda}[\tilde{\pi}^{\perp\perp}\lambda^{KL}(\nu\sinh\alpha-N^{\perp})
 +\tilde{\pi}^{KL}
N^{\perp}+l^{KL}N-l\lambda^{KL}\nu\cosh\alpha \nonumber \\
& +\nu^{M}\lambda^{KL}\alpha_{,M}-N_{,k}m^{k}\lambda^{KL}] 
- 8 \pi G T^{KL}_{s} = -\sqrt{\lambda}[\dot\alpha]\lambda^{KL}
\label{HK130} 
\end{align} 
\end{equation}
We get also a constraint, the so called \textit{$\alpha$-constraint} (129HK)
\begin{equation}
\begin{align}
\nu[\tilde{\pi}^{\perp\perp}\cosh\alpha-l\sinh\alpha]=0 
\label{alphaC}      
\end{align}
\end{equation}
and an equation of motion for $\lambda$ (131HK)
\begin{equation}
\begin{align}
\dot{\lambda}=2\lambda(-\nu\tilde{\pi}^{\perp\perp}\cosh\alpha+l\nu\sinh\alpha
+\nu^{K}_{||K})^{\pm}.
\label{lambda}
\end{align} 
\end{equation}
The trace of (\ref{HK130}) is an equation of motion for $[\alpha]$ (137HK),
\begin{equation}
\sqrt{\lambda}[\dot{\alpha}] = 4 \pi G T^{KL}\lambda_{KL}+\frac{1}{2}
\sqrt{\lambda}[\tilde{\pi}^{KL}\lambda_{KL}N^{\perp}-lN+2\nu^{M}\alpha_{,M}
-2N_{,k}m^{k}]
\end{equation}
while the trace-free part of the same (\ref{HK130}) is a constraint, 
the so called $\lambda$-\textit{constraint} (137.2HK)
\begin{equation}
\begin{align}
& -\sqrt{\lambda}[N^{\perp}(\tilde{\pi}^{KL}-\frac{1}{2}\tilde{\pi}^{MN}
\lambda_{MN}\lambda^{KL})+N(l^{KL}-\frac{1}{2}l\lambda^{KL})]= \nonumber \\
& = 8 \pi G (T^{KL}-\frac{1}{2}T^{MN}\lambda_{MN}\lambda^{KL})
\label{lambdaC}
\end{align}
\end{equation}
We have explicitly reported the calculations of the HK paper about 
$\delta\mathcal{H}_{118}$ and $\delta\mathcal{H}_{107}$, because we will start 
from here in order to see how they modify when the geometrical relations 
described in Section \ref{secAC} will be taken into account.\\ \\
We see that the constraints $\gamma$-C, $\alpha$-C, $\lambda$-C are 
\textit{non differentiable} on the phase space. In fact, $\gamma$-C, $\alpha$-C, 
$\lambda$-C are all calculated and defined on the shell $S \cap \Sigma$, 
therefore they can be smeared only by functions of \textit{two} variables.
But they contain volume quantities (namely $q_{kl}$, 
$\tilde{\pi}^{\perp\perp}=(\pi^{kl}m_{k}m_{l})/\sqrt{q}$, $l=q^{kl}m_{k|l}$,
$\tilde{\pi}^{KL}$), i.e. quantities that are (a priori) defined also outside 
the shell $S \cap \Sigma$. Poisson Brackets of these constraints would involve 
functional derivatives in respect to these volume variables. This can generate 
3-dimensional Dirac $\delta$-functions that cannot be re-absorbed via integration,
because we can integrate only on $S \cap \Sigma$ (that is, only in $d^{2}\eta$
and not in $d^{3}y$). So, if we want to have well defined Poisson Brackets, 
we need differentiable constraints and hamiltonian: therefore, we have to "eliminate" 
the \textit{singular} (non-differentiable) constraints.

\section{The reduction procedure}

We will see that the $\gamma$-C will be solved via a particular coordinate system.
To eliminate the singular constraints $\alpha$-C and $\lambda$-C, we adopt a 
reduction procedure due to Teitelboim and Henneaux. In the following, we rapidly 
sketch the "hamiltonian" version of this procedure.\\
If $\mathcal{H}=\mathcal{H}(x,y,z)$ is the hamiltonian of our system and
$\delta\mathcal{H}^{SYM}=A\delta x+B\delta y$ is the symplectic form, we can get 
equations of motion and constraints by comparing $\delta\mathcal{H}$ with
$\delta\mathcal{H}^{SYM}$
\begin{equation}
\begin{align}
\delta\mathcal{H}=\frac{\partial\mathcal{H}}{\partial x}\delta x +    
\frac{\partial\mathcal{H}}{\partial y}\delta y +
\frac{\partial\mathcal{H}}{\partial z}\delta z = 
A\delta x+B\delta y + 0 \cdot \delta z = \delta\mathcal{H}^{SYM}
\end{align}
\end{equation}

\[ \Rightarrow \left\{ \begin{array}{lll}
\frac{\partial\mathcal{H}}{\partial x}(x,y,z)=A  \\ \\
\frac{\partial\mathcal{H}}{\partial y}(x,y,z)=B  \\ \\
\frac{\partial\mathcal{H}}{\partial z}(x,y,z)=g(x,y,z)=0
\end{array}
\right. \]
The last equation is a constraint. We solve the constraint in respect to the variable
whose variation generates it.
\begin{equation}
g(x,y,z)=0 \quad \Longrightarrow \quad z=\bar{z}(x,y)\quad s.t.\quad 
g(x,y,\bar{z}(x,y))\equiv 0.
\end{equation}
Then we substitute the solution $\bar{z}(x,y)$ back into the hamiltonian
$\mathcal{H}=\mathcal{H}(x,y,\bar{z})$ and in the symplectic form       
$\delta\mathcal{H}^{SYM}=A\delta x+B\delta y + 0\cdot\delta\bar{z}$.
The variations now become
\begin{equation}
\delta\mathcal{H}=\left(\frac{\partial\mathcal{H}}{\partial x} +    
\frac{\partial\mathcal{H}}{\partial \bar{z}}\frac{\partial\bar{z}}{\partial x}
\right)\delta x +
\left(\frac{\partial\mathcal{H}}{\partial y} +    
\frac{\partial\mathcal{H}}{\partial \bar{z}}\frac{\partial\bar{z}}{\partial y}
\right)\delta y 
= 
A\delta x+B\delta y + 0 \cdot \delta \bar{z} = \delta\mathcal{H}^{SYM}.  
\end{equation}
But
\begin{equation}
\frac{\partial\mathcal{H}}{\partial \bar{z}}(x,y,\bar{z})=g(x,y,\bar{z}(x,y))\equiv 0.
\end{equation}
Therefore the new equations are

\[ \left\{ \begin{array}{ll}
\frac{\partial\mathcal{H}}{\partial x}(x,y,\bar{z}(x,y))=A(x,y,\bar{z})  \\ \\
\frac{\partial\mathcal{H}}{\partial y}(x,y,\bar{z}(x,y))=B(x,y,\bar{z})  \\ \\
\end{array}
\right. \]
and the constraint $g=0$ has completely disappeared (i.e. it is identically 
satisfied).  

\section{Solution of the $\gamma$ Constraint}

\label{secGammaC}
We want now to find the variables in which the $\gamma$-C is automatically 
satisfied, i.e. we are going to build the theoretical set-up which consistently 
implements the $\gamma$-C. \\  \\
\underline{Gaussian Coordinates}: Firstly we observe that we can build a set-up
in which the $\gamma$-C is satisfied \textit{by construction}, if we use gaussian 
coordinates in $M$ based on $\Sigma$. In fact, with coordinates of the kind

\begin{tabbing}
$\xi^{0}=x^{0}(y^{1}, y^{2}, y^{3})$ \\
$\xi^{1}=x^{1}$ \\ 
$\xi^{2}=x^{2}$ \\    
$x^{3}=x^{3}_{G}$ (=geodesic distance normal to $\Sigma$)
\end{tabbing}
we can write the line element in the following way:

\begin{equation}
ds^{2}_{M}=ds^{2}_{\Sigma}+(dx_{G})^{2}=\gamma_{\alpha\beta}
d\xi^{\alpha}d\xi^{\beta}+(dx_{G})^{2}
\label{219}
\end{equation}
where $\gamma_{\alpha\beta}$ is a given metric on the $x^{3}=const.$ 
surfaces whose $\Sigma$ is an example. From here we obtain the form of 
$g_{\mu\nu}$

\[ g_{\mu\nu} =\left[
\begin{array}{llll}
\; & \; & \; & 0  \\    
\; & \gamma_{\alpha\beta}^{(x^{3})} & \; & 0 \\
\; & \; & \; & 0 \\
0 & \;\;\; 0 & 0 & 1
\end{array} 
\right]   \]
This expression of $g_{\mu\nu}$ is valid all 
over $M$ (in fact we could say, in other words, that the conditions to be 
imposed on $g_{\mu\nu}$ to reduce it to that form are 4, and therefore 
no given geometry is \textit{a priori} fixed).
If we assign the metrics $\gamma_{\alpha\beta}^{(x^{3})}$ 
on $x^{3}=const.$
surfaces and we use gaussian coordinates, the metrics 
$\gamma_{\alpha\beta}^{(x^{3})}$ determine $g_{\mu\nu}$ all over M.
Besides, by construction, the $g_{\mu\nu}$ obtained in such way 
automatically satisfy the $\gamma$-C:

\begin{equation}
\begin{align}
& (g_{\mu\nu}e^{\mu}_{\alpha}e^{\nu}_{\beta})^{+}=(g_{\mu\nu}\delta^{\mu}_{\alpha}
\delta^{\nu}_{\beta})^{+}=g_{\alpha\beta}^{+}=\gamma_{\alpha\beta} \nonumber \\
& (g_{\mu\nu}e^{\mu}_{\alpha}e^{\nu}_{\beta})^{-}=(g_{\mu\nu}\delta^{\mu}_{\alpha}
\delta^{\nu}_{\beta})^{-}=g_{\alpha\beta}^{-}=\gamma_{\alpha\beta} 
\label{220}
\end{align}
\end{equation}
Indeed, because we are interested only in the form of the metric $g_{\mu\nu}(x)$
when $x \rightarrow \Sigma$, it is sufficient for our scope, to assign simply the
metric $\gamma_{\alpha\beta}$ on $\Sigma$. The (\ref{219}) will be then valid only
in a neighbourhood of $\Sigma$ (not all over $M$), but this is sufficient to
affirm the (\ref{220}).    
So, the use of gaussian coordinates permit to us to satisfy the $\gamma$-C. 
However, we will not adopt gaussian coordinates in the following, because they
imply a too much strong reduction, as we see from the structure of the $g_{\mu\nu}$
matrix. In other words, there are not enough variables to be varied: in particular,
some of the lagrangian multipliers $N$, $N^{k}$ are arbitrarily set equal to zero
and this would prevent us from writing the correct $3+1$ decomposition of the hamiltonian.       
\\ \\
\underline{Adapted Coordinates}: We now will see how it is possible to determine
a metric $g_{\mu\nu}$ in M from given metrics $\gamma_{\alpha\beta}$
on $x^{3}=const.$ surfaces, also if we work with the more general frame of
adapted coordinates. The obtained $g_{\mu\nu}$ will automatically satisfy the  
$\gamma$-C. The coordinates are adapted as before, i.e. in the usual HK paper 
sense
\begin{tabbing}
$x^{0}=t$ on $S_{t}$ \\
$x^{3}=0$ on $\Sigma$ \\ 
$x^{k}|_{S_{t}}=y^{k}$; $x^{\alpha}|_{\Sigma}=\xi^{\alpha}$  \\    
$x^{A}|_{S\cap\Sigma}=\eta^{A}$
\end{tabbing}
We can allow here a more general situation than the previous one (in gaussian 
coordinates): the metric can also not be continuos across $\Sigma$. At this 
point, fixed a metric $\gamma_{\alpha\beta}$ on the $x^{3}=const.$ surfaces,
we can write for the line element in $V^{\pm}$
\begin{equation}
ds^{2}_{M}=\gamma_{\alpha\beta}(d\xi^{\alpha}+M^{\alpha}dx^{3})
(d\xi^{\beta}+M^{\beta} dx^{3}) + (Mdx^{3})^{2}
\end{equation}
From here we deduce that the metric $g_{\mu\nu}$ in  $V^{\pm}$ is
\[ (g_{\mu\nu})^{\pm} =\left[
\begin{array}{llll}
\; & \; & \; & \;  \\    
\; & \gamma_{\alpha\beta}^{(x^{3})} & \; & \quad \quad (M_{\alpha})^{\pm} \\
\; & \; & \; & \; \\
\; & (M_{\beta})^{\pm} & \; & (M_{\alpha}M^{\alpha}+M^{2})^{\pm}
\end{array} 
\right]   \]
The indexes $\pm$ have been written because the two Cauchy surfaces 
$S^{\pm}$ where the coordinate lines $x^{3}$ lie (in $V^{\pm}$), 
have different normals: $n_{\mu}^{+}\neq n_{\mu}^{-}$.
Therefore the decomposition has been done with functions $M_{\alpha}$ that 
take different values in $V^{+}$ and $V^{-}$.
The metric $g_{\mu\nu}$ is, in general, not continuos at $\Sigma$, i.e. 
$(g_{\mu\nu})^{+}\neq(g_{\mu\nu})^{-}$ just because $(M_{\alpha})^{+}
\neq(M_{\alpha})^{-}$.
In other words we can allow jumps in the functions $M_{\alpha}$ and $M$ across
$\Sigma$. The metric built in this way obeys automatically the $\gamma$-C:

\begin{equation}
\begin{align}
& (g_{\mu\nu}e^{\mu}_{\alpha}e^{\nu}_{\beta})^{+}=(g_{\mu\nu}\delta^{\mu}_{\alpha}
\delta^{\nu}_{\beta})^{+}=g_{\alpha\beta}^{+}=\gamma_{\alpha\beta} \nonumber \\
& (g_{\mu\nu}e^{\mu}_{\alpha}e^{\nu}_{\beta})^{-}=(g_{\mu\nu}\delta^{\mu}_{\alpha}
\delta^{\nu}_{\beta})^{-}=g_{\alpha\beta}^{-}=\gamma_{\alpha\beta} 
\end{align}
\end{equation}
Note that the \textit{continuity of the whole metric} $g_{\mu\nu}$ is a request stronger 
than the simple $\gamma$-C, that demands, on the contrary, only the continuity of
\textit{a part} of $g_{\mu\nu}$.     
We can project the (\ref{gammaC}) on $S\cap\Sigma$ obtaining so the $2+1$ version 
of the $\gamma$-C. The induced metric on the Cauchy surfaces is 
$q_{kl}^{\pm}:=(g_{\mu\nu}e^{\mu}_{k}e^{\nu}_{l})^{\pm}$; the induced metric on 
$S\cap\Sigma$ from $\gamma_{\alpha\beta}$ on $\Sigma$ is $\lambda_{AB}=
\gamma_{\alpha\beta}e_{A}^{\alpha}e_{B}^{\beta}$; for the $\gamma$-C we can write:
$\gamma_{\alpha\beta}=(g_{\mu\nu}e^{\mu}_{\alpha}e^{\nu}_{\beta})^{\pm}$. 
Therefore we can deduce from here:
\begin{equation}
\begin{align}
\lambda_{AB} & =\gamma_{\alpha\beta}e_{A}^{\alpha}e_{B}^{\beta}=
(g_{\mu\nu}e^{\mu}_{\alpha}e^{\nu}_{\beta})^{\pm}e_{A}^{\alpha}e_{B}^{\beta}=
(g_{\mu\nu}e^{\mu}_{A}e^{\nu}_{B})^{\pm} \nonumber \\
& =
(g_{\mu\nu}e^{\mu}_{k}e^{\nu}_{l})^{\pm}e^{k}_{A}e^{l}_{B}=
q_{kl}^{\pm}e^{k}_{A}e^{l}_{B}.
\end{align}
\end{equation}
The 2+1 version of the $\gamma$-C is therefore
\begin{equation}
\lambda_{AB}=(q_{kl}e^{k}_{A}e^{l}_{B})^{\pm}.
\end{equation}
We note also here that the 2+1 version does not imply the continuity of the 
3-metric $q_{kl}$.
The $\gamma$-C is therefore automatically satisfied in adapted coordinates.

\section{Connection between the $\alpha$ Constraint and the Dynamics}

\label{dyn}
Observe that the relation $\lambda_{AB}(t)=(q_{kl}(t)e^{k}_{A}e^{l}_{B})^{\pm}$
holds for any value of the temporal coordinate $t$. On the other hand, repeating 
the calculations in adapted coordinates already done in section \ref{secH118}, we can 
obtain, from the equation $\delta\mathcal{H}_{118}=\delta\mathcal{H}^{SYM}_{107}$,
the equation of motion $\dot{q}_{kl}=b_{kl}$, true outside $\Sigma$ at any time. 
Then, from the two equations

\begin{equation}
\begin{align}
\lambda_{AB} & =(q_{kl}e^{k}_{A}e^{l}_{B})^{\pm} \nonumber \\
\dot{q}_{kl} & =b_{kl}  
\end{align}
\end{equation}
we get (the expression for $b_{kl}$ is given in ~\cite{HK})
\begin{equation}
\begin{align}
\dot{\lambda}_{AB} & =\{\dot{q}_{kl}e^{k}_{A}e^{l}_{B}\}^{\pm}=
\{b_{kl}e^{k}_{A}e^{l}_{B}\}^{\pm}=
\{(\frac{N}{\sqrt{q}}(2\pi_{kl}-\pi q_{kl})+(N_{k|l}+N_{l|k}))
e^{k}_{A}e^{l}_{B}\}^{\pm} \nonumber \\
& =
\{N(2\tilde{\pi}_{AB}-\tilde{\pi}\lambda_{AB})+(N_{A||B}
+N_{B||A}+2l_{AB}N^{\perp})\}^{\pm} \label{132HK}
\end{align}
\end{equation}
If now we remind that $(N_{A})^{+}=(N_{A})^{-}$ 
(because $N_{A}=\textbf{e}_{0}\cdot\textbf{e}_{A}$), we can say that
\begin{equation}
\{... \;\; ...\}^{\pm}=\dot{\lambda}_{AB} \quad \Rightarrow \quad [... \;\; ...]=0
\nonumber
\end{equation}
i.e.
\begin{equation}
[N(2\tilde{\pi}_{AB}-\tilde{\pi}\lambda_{AB})+2l_{AB}N^{\perp}]=0
\end{equation}
and this holds at any time.\\
Taking the trace, we get
\begin{equation}
\begin{align}  
0 & =[N(2\tilde{\pi}_{AB}\lambda^{AB}-\tilde{\pi}\lambda_{AB}\lambda^{AB})
+2l_{AB}\lambda^{AB}N^{\perp}] \nonumber \\
& =
[N(2(\tilde{\pi}-\tilde{\pi}^{\perp\perp})-2\tilde{\pi})+2lN^{\perp}] \nonumber \\
& =
2[-N\tilde{\pi}^{\perp\perp}+N^{\perp}l]
\end{align}
\end{equation}
and this is the $\alpha$-C. As we see, we have obtained it not from a direct 
comparison between the variation of the hamiltonian and the symplectic form, 
but from the equation 
of motion and from the time derivative of the 2+1 version of the $\gamma$-C.\\
Moreover, if we take the trace of (\ref{132HK}) we get
\begin{equation}
\begin{align} 
\dot{\lambda}_{AB}\lambda^{AB} & =\{N(2\tilde{\pi}_{AB}\lambda^{AB}-\tilde{\pi}
\lambda_{AB}\lambda^{AB})+(N_{A||B}+N_{B||A}+2l_{AB}N^{\perp})\lambda^{AB}\}^{\pm} \nonumber \\
& =2\{-N\tilde{\pi}^{\perp\perp}+(N_{A}\lambda^{AB})_{||B}+lN^{\perp}\}^{\pm} \nonumber \\
& =2\{-N\tilde{\pi}^{\perp\perp}+N^{\perp}l+\nu^{B}_{||B}\}^{\pm}
\end{align}
\end{equation}   
(in the last step we have used the fact that $N_{A}=\nu_{A}$).
The last one is the (131HK), an equation of motion for $\lambda$. 
Also this equation holds at any time in adapted coordinates.
But these equations, at this stage, are only equations of motion, 
i.e. they hold along particular curves of the phase space.
They are not identities on the whole phase space. Therefore they cannot be used to 
reduce the phase space variables and to semplify the successive calculations. 
In other words, the $\alpha$-C still remains to be solved.

\section{Implementing the consequences of adapted coordinates}

The choice of adapted coordinates has the following consequences:\\
a) \[ \left\{ \begin{array}{ll}
\nu\sinh\alpha_{\pm}-N^{\perp}_{\pm}=0 \\ \\
\nu\cosh\alpha_{\pm}-N_{\pm}=0
\end{array}
\right. \quad \left\{ \begin{array}{ll}
\nu_{M}=\textbf{e}_{0} \cdot \textbf{e}_{M}=N_{M}\\ \\
\nu=\sqrt{(N_{\pm})^{2}-(N_{\pm}^{\perp})^{2}}
\end{array}
\right. \] 
(see section \ref{secAC} for a pure geometrical derivation).\\ \\
b) Validity of the $\gamma$-C: $(g_{\mu\nu}e^{\mu}_{\alpha}
e^{\nu}_{\beta})^{\pm}= g_{\alpha\beta}^{\pm}= \gamma_{\alpha\beta}$ \\
and of its 2+1 version: $(q_{kl}e^{k}_{A}e^{l}_{B})^{\pm}=q_{AB}|_{S\cap\Sigma}=
\lambda_{AB}$\\
(see section \ref{secGammaC} for a derivation from geometry).\\ \\
In section \ref{secH118}, as in the original HK paper, 
the calculation were done without taking into account all the 
consequences derived from the choice of adapted coordinates.
The variations were calculated by thinking all the variables 
as if they were reciprocally independent.
Here, on the contrary, we will take into account from the beginning the important
identities listed before and we will use them to redefine the phase space.\\ \\ 
\underline{New phase space}: The relations listed before imply a redefinition of the
variables describing the phase space. In particular $\nu$, $\sinh\alpha$, $\cosh\alpha$
disappear and they are substituted by $N$, $N^{\perp}$. Also the variable $\lambda_{AB}$
disappears as independent variable: it coincides with $q_{kl}$ at $\Sigma$.
From the identities $a)$ we get
\begin{equation}
\alpha_{\pm}=\tanh^{-1}\frac{N^{\perp}_{\pm}}{N_{\pm}}
\label{ArcTh}
\end{equation}     
The new phase space is therefore described by the following canonical variables

\( \begin{array}{cll}
\hspace*{3cm}-\frac{1}{16 \pi G}\;q_{kl} & \longleftrightarrow & \pi^{kl} \\ \\
\hspace*{3cm}-\frac{1}{8 \pi G}\sqrt{\lambda} & \longleftrightarrow & 
[\tanh^{-1}\frac{N^{\perp}}{N}] \\ \\
\hspace*{3cm}z^{A} & \longleftrightarrow & p_{A}
\end{array} \) \\ \\
Hereafter we will present the new form assumed by the hamiltonian, by the symplectic form,
by the new equations and constraints.\\
The strategy followed in the calculus of these new expressions is as much direct as 
possible: we consider the expressions of $\mathcal{H}_{118}$, 
$\delta\mathcal{H}_{118}$, $\delta\mathcal{H}_{107}^{SYM}$ as given in HK paper and 
we transform them by the known substitutions already listed. Other substitutions 
will be presented during the calculations in order to eliminate "old" parameters.
In the following $\lambda_{AB}$ is simply an abreviation for $q_{AB}|_{S\cap\Sigma}$.\\
\underline{New hamiltonian}: 
For the identities previously stated we see that
\begin{equation}
[\alpha_{,K}]=(\alpha^{+}-\alpha^{-})_{,K}=[(\tanh^{-1}\frac{N^{\perp}}{N})_{,K}]
\end{equation}
and 
\begin{equation}
\nu^{K}=\lambda^{KL}\nu_{L}=\lambda^{KL}N_{L}.
\label{nukappa}
\end{equation}
Starting from the expression (\ref{H118}) of the old hamiltonian
$\mathcal{H}_{118}$ we can write the expression for the new hamiltonian
\begin{equation}
\begin{align} 
\mathcal{H}_{118}^{NEW} & =\frac{1}{16 \pi G} \int_{S^{-}\cup S^{+}}d^{3}y\{NC+N^{k}C_{k}\} \nonumber \\
& -\frac{1}{8 \pi G} \int_{S\cap\Sigma}d^{2}\eta\sqrt{\lambda}
([\tilde{\pi}^{\perp\perp}N^{\perp}-lN]+\lambda^{KL}N_{L}[\tilde{\pi}^{\perp}_{K}+
(\tanh^{-1}\frac{N^{\perp}}{N})_{,K}]) \nonumber \\
& -\int_{S\cap\Sigma}d^{2}\eta T^{0}_{s\;0}
\end{align}
\end{equation}
\underline{New symplectic form}: Reminding the expression of the "old" 
symplectic form, we can now write
\begin{equation}
\begin{align}
\delta\mathcal{H}^{SYM}_{107(NEW)} & = \frac{1}{16 \pi G}\int_{S}d^{3}y(-\dot{\pi}^{kl}
\delta q_{kl}+\dot{q}_{kl}\delta\pi^{kl}) \nonumber \\
& +\frac{1}{16 \pi G}\int_{S\cap\Sigma}d^{2}
\eta\sqrt{\lambda}\lambda^{KL}(-\dot{[\tanh^{-1}\frac{N^{\perp}}{N}]}\delta\lambda_{KL}
+\dot{\lambda}_{KL}\delta[\tanh^{-1}\frac{N^{\perp}}{N}]) \nonumber \\ 
& -\int_{S\cap\Sigma}d^{2}\eta(\dot{p}_{A}\delta
z^{A}-\dot{z}^{A}\delta p_{A}).
\end{align}
\end{equation}
From here we read the new phase space variables that are those showed in the last but one 
paragraph.\\
\underline{New variation of $\mathcal{H}_{118}$}: 
To calculate $\delta\mathcal{H}_{118}^{NEW}$, we should keep in mind the "old" 
variation $\delta\mathcal{H}_{118}$ and the validity of the chain-rule for the 
variational calculus. Besides to the usual identities 
(i.e. $\lambda_{KL}=q_{KL}|_{S\cap\Sigma}$, $\nu^{K}=\lambda^{KL}N_{L}$, 
$\alpha_{\pm}=\tanh^{-1}(N^{\perp}_{\pm}/N_{\pm})$), 
we need also the expression of $\nu$
as a function of $N_{\pm}$, $N^{\perp}_{\pm}$ (see relations in a)).
Therefore the variation $\delta\mathcal{H}_{118}^{NEW}$ can be written as
\begin{equation}
\begin{align}
\delta\mathcal{H}_{118}^{NEW} & =\frac{1}{16 \pi G}\int_{S}d^{3}y(C_{k}\delta N^{k}+
C\delta N + a^{kl}\delta q_{kl} + b_{kl}\delta\pi^{kl}) \nonumber \\
& +\frac{1}{16 \pi G}
\int_{S\cap\Sigma}d^{2}\eta[2REST-\sqrt{\lambda}B^{k}m_{k}] \nonumber \\
& -\frac{1}{8 \pi G}\int_{S\cap\Sigma}d^{2}\eta\frac{\sqrt{\lambda}}
{\sqrt{|\gamma|}}
\{\left([Q^{\perp\perp}]-8\pi G T^{\perp\perp}_{s}\right)
\delta\left(\sqrt{(N_{\pm})^{2}-(N_{\pm}^{\perp})^{2}}\right) \nonumber \\
& +\left([Q^{\perp}_{K}]-8 \pi GT^{\perp}_{s\;K}\right)\delta(\lambda^{KL}N_{L})\}
- \int_{S\cap\Sigma}d^{2}\eta\left\{\frac{1}{2}T^{KL}_{s}\delta\lambda_{KL}\right\} \nonumber \\
& - \int_{S\cap\Sigma}d^{2}\eta\left\{\left(\frac{\partial T^{0}_{s\;0}}
{\partial z^{A}}-\frac{\partial}{\partial\eta^{M}}\frac{\partial T^{0}_{s\;0}}
{\partial z^{A}_{M}}\right)\delta z^{A}+ \frac{\partial T^{0}_{s\;0}}{\partial 
p_{A}}\delta p_{A}\right\}
\label{varH118new}
\end{align}
\end{equation}
\textit{\underline{Important note}}: We note that the $5$ variables $\nu$, $\nu^{K}$, 
$\alpha^{\pm}$ do not appear in 
$\delta\mathcal{H}_{118}^{NEW}$: they have been replaced by 
the $6$ variables $N_{\pm}$, $N^{\perp}_{\pm}$, $N_{1}$, $N_{2}$.
Because of the relation
\begin{equation}
(N_{+})^{2}-(N_{+}^{\perp})^{2}=(N_{-})^{2}-(N_{-}^{\perp})^{2}
\end{equation}
the free variables of the last group are still $5$.
Moreover here it is understood that $\lambda_{KL}=q_{KL}|_{S\cap\Sigma}$ (the $2+1$
form of the $\gamma$-C). We see that the $\gamma$-C does not produce an effective
reduction in the number of free variables entering in the new equations, but 
rather a re-naming of them, and a more correct identification of the "true" phase 
space variables.\\
\underline{Old and new equations}: We compare now
\begin{equation}
\delta\mathcal{H}_{118}^{NEW}=\delta\mathcal{H}^{SYM}_{107(NEW)}
\end{equation}
Apart from the "REST" terms, which will be considered in a moment, we obtain in 
the usual way the "new" canonical equations

\[ \left\{ \begin{array}{ll}
C_{k}=0 \\ \\
C=0
\end{array}
\right. \quad \left\{ \begin{array}{ll}
\dot{q}_{kl}=b_{kl} \\ \\
\dot{\pi}^{kl}=-a^{kl}
\end{array}
\right.  \quad \left\{ \begin{array}{ll} 
[Q^{\perp\perp}]=8 \pi G T^{\perp\perp}_{s} \\ \\

[Q^{\perp}_{K}]= 8 \pi G T^{\perp}_{s\;K}
\end{array}
\right.    \quad \left\{ \begin{array}{ll}
\dot{p}_{A}=\frac{\partial T^{0}_{s\;0}}
{\partial z^{A}}-\frac{\partial}{\partial\eta^{M}}\frac{\partial T^{0}_{s\;0}}
{\partial z^{A}_{M}} \\ \\
-\dot{z}^{A}=\frac{\partial T^{0}_{s\;0}}{\partial p_{A}}
\end{array}
\right.    \] \\ \\
which are clearly unchanged in form.
The remaining "REST" terms to be compared are
\begin{equation}
\begin{align}
& \int_{S\cap\Sigma}d^{2}\eta[2REST-\sqrt{\lambda}B^{k}m_{k}]
-\int_{S\cap\Sigma}d^{2}\eta \{8 \pi G T^{KL}_{s}\delta\lambda_{KL}\} = \nonumber \\
& \int_{S\cap\Sigma}d^{2}
\eta\sqrt{\lambda}\lambda^{KL}(-\dot{[\tanh^{-1}\frac{N^{\perp}}{N}]}\delta\lambda_{KL}
+\dot{\lambda}_{KL}\delta[\tanh^{-1}\frac{N^{\perp}}{N}]).
\end{align}
\end{equation}
In force of the identities a), b), (\ref{ArcTh}), (\ref{nukappa})
and $[A \cdot B]=A^{\pm}[B]+[A]B^{\mp}$, the last equation becomes
\begin{equation}
\begin{align}
 & \int_{S\cap\Sigma}d^{2}\eta(- \sqrt{\lambda}[\tilde{\pi}^{KL}
N^{\perp}+l^{KL}N-l\lambda^{KL}N+\lambda^{MR}N_{R}\lambda^{KL}(\tanh^{-1}\frac{N^{\perp}}{N})_{,M} \nonumber \\
& -N_{,k}m^{k}\lambda^{KL}]\delta\lambda_{KL}
+2\sqrt{\lambda}[-\tilde{\pi}^{\perp\perp}N+lN^{\perp}
+(\lambda^{KL}N_{L})_{||K}]\delta(\tanh^{-1}\frac{N^{\perp}_{\pm}}{N_{\pm}}) \nonumber \\
& +2\sqrt{\lambda}(-\tilde{\pi}^{\perp\perp}N+lN^{\perp}
+(\lambda^{KL}N_{L})_{||K})^{\pm}\delta[\tanh^{-1}\frac{N^{\perp}}{N}])
-\int_{S\cap\Sigma}d^{2}\eta(8 \pi G T^{KL}_{s}\delta\lambda_{KL})= \nonumber \\
& \int_{S\cap\Sigma}d^{2}
\eta\sqrt{\lambda}\lambda^{KL}(-\dot{[\tanh^{-1}\frac{N^{\perp}}{N}]}\delta\lambda_{KL}
+\dot{\lambda}_{KL}\delta[\tanh^{-1}\frac{N^{\perp}}{N}])
\label{var1.42}
\end{align}
\end{equation}  
From the comparison we finally obtain the equations:\\
1)- The new form of the $\alpha$-C
\begin{equation}
[-\tilde{\pi}^{\perp\perp}N+lN^{\perp}]=0
\label{newalphaC}
\end{equation}
2)- The new form of the equation of motion 131HK
\begin{equation}
2\sqrt{\lambda}(-\tilde{\pi}^{\perp\perp}N+lN^{\perp}
+(\lambda^{KL}N_{L})_{||K})^{\pm}=\sqrt{\lambda}\lambda^{KL}\dot{\lambda_{KL}}
\label{new131HK}
\end{equation}
3)- The new form of the equation 130HK
\begin{equation}
\begin{align}
& - \sqrt{\lambda}[\tilde{\pi}^{KL}N^{\perp}+l^{KL}N-l\lambda^{KL}N
+\lambda^{MR}N_{R}\lambda^{KL}(\tanh^{-1}\frac{N^{\perp}}{N})_{,M}
-N_{,k}m^{k}\lambda^{KL}] \nonumber \\
& -8 \pi GT^{KL}_{s}=\sqrt{\lambda}\lambda^{KL}(-\dot{[\tanh^{-1}\frac{N^{\perp}}{N}]}).
\label{new130HK}
\end{align}
\end{equation}
The trace of the last equation is the equation of motion for the variable
$[\tanh^{-1}(N^{\perp}/N)]$
\begin{equation}
\begin{align}
\sqrt{\lambda}[\dot{\tanh^{-1}\frac{N^{\perp}}{N}}] = & 4 \pi G T^{KL}\lambda_{KL}+\frac{1}{2}
\sqrt{\lambda}[\tilde{\pi}^{KL}\lambda_{KL}N^{\perp}-lN \nonumber \\
& +2(\lambda^{MR}N_{R})(\tanh^{-1}\frac{N^{\perp}}{N})_{,M}
-2N_{,k}m^{k}]
\end{align}
\end{equation}
while the trace-free part of the same equation is again the "old" 
$\lambda$ constraint
\begin{equation}
\begin{align}
& -\sqrt{\lambda}[N^{\perp}(\tilde{\pi}^{KL}-\frac{1}{2}\tilde{\pi}^{MN}
\lambda_{MN}\lambda^{KL})+N(l^{KL}-\frac{1}{2}l\lambda^{KL})]= \nonumber \\
& = 8 \pi G (T^{KL}-\frac{1}{2}T^{MN}\lambda_{MN}\lambda^{KL})
\end{align}
\end{equation}
which remains therefore unchanged in form, although now, of course, 
$\lambda_{KL}=q_{KL}|_{S\cap\Sigma}$.\\
We have seen in this way what are the consequences of the choice of 
adapted coordinates: solution of the $\gamma$-C, the changes in the symplectic 
form (phase space),
the new hamiltonian, the new form of the equations of motion and constraints.
Two constraints now remain to be solved, namely the $\alpha$-C and the $\lambda$-C.
They will be dealt with in the next chapters.

\section{Commutativity of the substitution procedure}

In this section we explicitly show the commutativity of the substitution procedure 
adopted in the last section. In more mathematical words, we will prove the commutativity
of the diagram 
\[ 
\begin{array}{lll}
\mathcal{H} & \stackrel{subst.}{\longrightarrow} & \mathcal{H}' \\    
\delta\downarrow  & \quad \quad & \downarrow\delta \\ 
eqs. & \stackrel{subst.}{\longrightarrow} & eqs.' 
\end{array}  
\quad \quad \quad (*) \]
In the last section we have essentially explored the upper-right sides of the 
diagram. We have shown how the new hamiltonian $\mathcal{H}_{118}^{NEW}$ can be 
obtained from the old one $\mathcal{H}_{118}$ via the substitutions dictated 
by the geometrical properties a), b) (and using also the 
property $\nu^{K}=\lambda^{KL}N_{L}$ derived from the geometrical identity 
$\nu_{M}=N_{M}$). We have described the new phase space and the new symplectic 
form. Then we have varied $\mathcal{H}_{118}^{NEW}$ and we have obtained,
besides the usual canonical equations, an equation of motion for 
$[\tanh^{-1}(N^{\perp}/N)]$ and two constraints, one corresponding to the old 
$\alpha$-C (\ref{alphaC}) and the other coinciding with the old $\lambda$-C
(\ref{lambdaC}), of course via $\lambda_{AB}=q_{AB}|_{\Sigma}$.\\
Now let's explore the left-lower part of the diagram. Consider the old hamiltonian
$\mathcal{H}_{118}$: with the variation (\ref{varH118}) and the comparison
$\delta\mathcal{H}_{118}=\delta\mathcal{H}_{107}^{SYM}$ we obtain the canonical 
equations. We obtain also the equations  

\[ (1)\left\{ \begin{array}{ll}
\nu\sinh\alpha_{\pm}=N^{\perp}_{\pm} \\ \\
\nu\cosh\alpha_{\pm}=N_{\pm}
\end{array}
\right. \] 
and these equations are obtained in form of constraints, dictated by 
the hamiltonian, i.e. by the dynamics. On the contrary, we remember, 
they appear in the first approach (upper-left sides of the diagram), 
as properties dictated by the geometry, preceding the dynamics.  
Besides we get the equations
\begin{equation}
[\tilde{\pi}^{\perp\perp}\cosh\alpha-l\sinh\alpha]=0 \quad \quad \quad \quad \quad \quad (\alpha-C)
\nonumber
\end{equation}
\begin{equation}
2\sqrt{\lambda}(-\nu\tilde{\pi}^{\perp\perp}\cosh\alpha+l\nu\sinh\alpha
+\nu^{K}_{||K})^{\pm} = \sqrt{\lambda}\lambda^{KL}\dot{\lambda}_{KL} \quad \quad \quad \quad (131HK)
\nonumber
\end{equation}
\begin{equation}
\begin{align}
& -\sqrt{\lambda}[\tilde{\pi}^{\perp\perp}\lambda^{KL}(\nu\sinh\alpha-N^{\perp})
 +\tilde{\pi}^{KL}
N^{\perp}+l^{KL}N-l\lambda^{KL}\nu\cosh\alpha \nonumber \\
& +\nu^{M}\lambda^{KL}\alpha_{,M}
-N_{,k}m^{k}\lambda^{KL}] - 8 \pi G T^{KL}_{s} = -\sqrt{\lambda}[\dot\alpha]\lambda^{KL}
\quad \quad \quad \quad (130HK)
\nonumber
\end{align} 
\end{equation}
From the last equation, the equation of motion 137HK (by trace) and the $\lambda$-C 
(by trace-free) come out.\\
We see at a glance that the substitutions a), b), $\nu^{K}=\lambda^{KL}N_{L}$
of the last section make the preceding equations to coincide with those 
obtained in the last section. Precisely:\\
The equations (1) become identities (in force of a));\\
The equation $\alpha$-C becomes $[-\tilde{\pi}^{\perp\perp}N+lN^{\perp}]=0$, (\ref{newalphaC});\\
The equation 131HK becomes the "new" 131HK, (\ref{new131HK});\\
The equation 130HK becomes the "new" 130HK, (\ref{new130HK}).\\
We see that all these equations coincide with those obtained in the preceding 
section. The commutativity of the diagram is therefore fully proved.
%
%
%
%
%
%
\chapter{Explicit solution of the $\alpha$ Constraint}
In this chapter we deal with the explicit solution of the $\alpha$-C,
in order to get rid of this \textit{singular} ($\equiv$ non differentiable) 
constraint. The form of the $\alpha$-C left to us from the calculations 
of the preceding section is
\begin{equation}
[N\tilde{\pi}^{\perp \perp}-lN^{\perp}]=0
\end{equation}
We can use also auxiliary variables in the calculations, 
provided that the relations between auxiliary variables and phase space 
variables be clearly stated, and the variational calculus rules
followed during the various steps. 
Reminding the last chapter, we can summarize
\[ \begin{array}{ccc}
AUXILIARY & \quad & PHASE \; SPACE \\
VARIABLES & \quad & VARIABLES \\ \\
\alpha_{+} & \quad & N_{+}^{\perp}|_{\Sigma} \\ \\
\alpha_{-} & \quad & N_{-}^{\perp}|_{\Sigma} \\ \\
\nu & \quad & N_{+}|_{\Sigma}, \;  N_{-}|_{\Sigma} \\ \\
\lambda_{AB} & \quad & q_{AB}|_{\Sigma} \\ \\
\nu^{K} & \quad & N_{L}|_{\Sigma}
\end{array} \]
(where $\lambda^{KL}=q^{KL}|_{\Sigma}$) and the relations are
\[ \begin{array}{lll}
N^{\perp}_{\pm}=\nu\sinh\alpha_{\pm} & \quad & \lambda_{KL}=q_{KL}|_{\Sigma} \\ \\
N_{\pm}=\nu\cosh\alpha_{\pm} & \quad & \nu^{K}=\lambda^{KL}N_{L}
\end{array} \]
The variables $N^{\perp}_{\pm}$, $N_{\pm}$ are therefore linked by the 
relation
\begin{equation}
(N_{+})^{2}-(N_{+}^{\perp})^{2}=(N_{-})^{2}-(N_{-}^{\perp})^{2}
\end{equation}
Hence the 3 free variables $\alpha_{+}$, $\alpha_{-}$, $\nu$ correspond
to the 3 free variables $N_{+}^{\perp}$, $N_{-}^{\perp}$, $N_{-}$.
Further, we will see that the solution of the $\alpha$-C will give us one more 
relation between $N^{\perp}_{\pm}$, $N_{\pm}$.
In order to do some calculations, we use now auxilary variables and in the end 
we shall restaure the real phase space variables.     
We will prove that the $\alpha$-C can be obtained by varying
$\mathcal{H}_{118}$ in respect to $\bar{\alpha}=\frac{1}{2}(\alpha^{+}+\alpha^{-})$.
Then, following the Teitelboim-Henneaux method, we will solve the 
$\alpha$-C explicitly for $\bar{\alpha}$ and we will insert the solution back 
into the hamiltonian. In this way the $\alpha$-C will be radically eliminated 
from the equations generated by the final hamiltonian. 
Moreover, it will be showed, in a completely general way,
that this modified hamiltonian not only does not give rise anymore to the singular
$\alpha$-C, but still it continues to produce correctly the other equations of 
motion and the other constraints.   

\section{Introduction of $\overline{\alpha}$ and [$\alpha$]}

We define two new variables 

\begin{equation}
\begin{align}
\bar{\alpha} & :=\frac{1}{2}(\alpha^{+}+\alpha^{-}) \nonumber \\
[\alpha] & :=\alpha^{+}-\alpha^{-}
\end{align}
\end{equation}
called respectively "mean value" and "jump" of $\alpha$. 
Of course also the inverse relations hold
\begin{equation}
\begin{align}
\alpha^{+}=\bar{\alpha}+\frac{1}{2}[\alpha] \nonumber \\
\alpha^{-}=\bar{\alpha}-\frac{1}{2}[\alpha]
\end{align}
\end{equation}
We note that not only $\alpha$ has a jump at $S\cap\Sigma$ but also 
$\pi^{kl}$, $m_{k}$, $l$ do have jumps there. 
Substituting $\alpha^{\pm}=\bar{\alpha}\pm\frac{1}{2}[\alpha]$ in the $\alpha$-sector
of the hamiltonian $\mathcal{H}_{118}$ (see relation (\ref{H118})), 
we obtain the hamiltonian as a functional of $\bar{\alpha}$, $[\alpha]$
\begin{multline}
\mathcal{H}_{118}(\bar{\alpha}, [\alpha]) =
-\frac{1}{8 \pi G} \int_{S\cap\Sigma}d^{2}\eta\sqrt{\lambda}
\{\nu((\tilde{\pi}^{\perp\perp}_{+}\sinh(\bar{\alpha}+\frac{1}{2}[\alpha]) \\
-l^{+}\cosh(\bar{\alpha}+\frac{1}{2}[\alpha])) 
-(\tilde{\pi}^{\perp\perp}_{-}\sinh(\bar{\alpha}-\frac{1}{2}[\alpha])
-l^{-}\cosh(\bar{\alpha}-\frac{1}{2}[\alpha]))) \\
+\nu^{K}
((\tilde{\pi}_{K}^{\perp +}+(\bar{\alpha}+\frac{1}{2}[\alpha])_{,K})
-(\tilde{\pi}_{K}^{\perp -}
+(\bar{\alpha}-\frac{1}{2}[\alpha])_{,K}))\}. 
\label{Halfa}
\end{multline}
We vary now $\mathcal{H}_{118}$ in respect to $\bar{\alpha}$, $[\alpha]$, 
both thought as variables. Then we compare $\delta\mathcal{H}_{118}$ with the
symplectic form $\delta\mathcal{H}_{107}^{SYM}$ (only with the $\alpha$-sector 
of it) in order to obtain equations of motion and constraints.

\begin{equation}
\begin{align}
\delta\mathcal{H}_{118}(/\delta\bar{\alpha}, \delta[\alpha]) =
& -\frac{1}{8 \pi G} \int_{S\cap\Sigma}d^{2}\eta\sqrt{\lambda}
\{\nu[\tilde{\pi}^{\perp\perp}\cosh\alpha-l\sinh\alpha]\delta\bar{\alpha} 
\nonumber \\ \nonumber \\
& +\overline{(\tilde{\pi}^{\perp\perp}\nu\cosh\alpha-l\nu\sinh\alpha)}\delta[\alpha]+
\nu^{K}(\delta[\alpha])_{,K}\}
\end{align}
\end{equation}
Observe that, because $\delta[\alpha]$ is a scalar, 
$(\delta[\alpha])_{,K}=(\delta[\alpha])_{||K}$. Therefore

\begin{equation}
\nu^{K}(\delta[\alpha])_{,K}=\nu^{K}(\delta[\alpha])_{||K} =
(\nu^{K}\delta[\alpha])_{||K}-\nu^{K}_{||K}\delta[\alpha] \nonumber
\end{equation}  
The term $(\nu^{K}\delta[\alpha])_{||K}$ represents a 2-divergence 
and it is integrated over $\int_{S\cap\Sigma}d^{2}\eta\sqrt{\lambda}$.
It therefore disappears from the final equations. Hence we consider only
$-\nu^{K}_{||K}\delta[\alpha]$.
Moreover $\nu_{A}=\mathbf{e}_{0}\cdot\mathbf{e}_{A}$, therefore
$(\nu_{A})^{+}=(\nu_{A})^{-}$ and thus $(\nu^{K}_{||K})^{+}=(\nu^{K}_{||K})^{-}$.
So, finally the last equation can be written as

\begin{equation}
\begin{align}
\delta\mathcal{H}_{118}=
& -\frac{1}{8 \pi G} \int_{S\cap\Sigma}d^{2}\eta\sqrt{\lambda}
\{\nu[\tilde{\pi}^{\perp\perp}\cosh\alpha-l\sinh\alpha]\delta\bar{\alpha} 
\nonumber \\ \nonumber \\
& +\overline{(\tilde{\pi}^{\perp\perp}\nu\cosh\alpha-l\nu\sinh\alpha- 
\nu^{K}_{||K})}\delta[\alpha]\} 
\end{align}
\end{equation}
This must be compared with the relevant terms of the symplectic form 
$\delta\mathcal{H}_{107}^{SYM}$
\begin{equation}
\delta\mathcal{H}_{107}^{SYM}(\alpha)= 
\frac{1}{16 \pi G} \int_{S\cap\Sigma}d^{2}\eta\sqrt{\lambda}
\left(\frac{\dot{\lambda}}{\lambda}\delta[\alpha]-
\dot{[\alpha]}\frac{\delta\lambda}{\lambda}\right).
\end{equation}

As we see, there are no terms in $\delta\bar{\alpha}$ in the symplectic form, 
and there is only one term in $\delta[\alpha]$. And because of the fact that
$\lambda$ is defined on $S\cap\Sigma$ and therefore doesn't change when $\Sigma$ 
is crossed, we can write $\lambda=\bar{\lambda}$ and $\lambda=\bar{\dot{\lambda}}$.
Hence the equations dictated by $\delta\mathcal{H}_{118}=\delta\mathcal{H}_{107}^{SYM}$ are

\begin{equation}
\begin{align}    
[\tilde{\pi}^{\perp\perp}\cosh\alpha-l\sinh\alpha] & =0 \nonumber \\ \nonumber \\
\overline{\tilde{\pi}^{\perp\perp}N-lN^{\perp}-\nu^{K}_{||K}} & =
-\frac{1}{2}\left(\frac{\bar{\dot{\lambda}}}{\lambda}\right)
\end{align}
\end{equation}
The first equation is the $\alpha$-C; the second is the "mean" of the equation 
of motion (131HK). 
We have shown in this way that the $\alpha$-C (129HK) follows from the variation 
of $\mathcal{H}_{118}$ with respect to $\bar{\alpha}$. 

\section{Solution of the $\alpha$-C in respect to $\bar{\alpha}$}

In this section we solve the $\alpha$-C in respect to $\bar{\alpha}$. 
Reminding that $\alpha^{\pm}=\bar{\alpha}\pm\frac{1}{2}[\alpha]$, in few passages
the $\alpha$-C can be written as
\begin{equation}
\mathcal{A}\sinh\bar{\alpha}+\mathcal{B}\cosh\bar{\alpha}=0
\end{equation}
where we have defined
\begin{equation}
\begin{align}
\mathcal{A}:= & \{2\overline{\tilde{\pi}^{\perp\perp}}\sinh\frac{[\alpha]}{2}-
[l]\cosh\frac{[\alpha]}{2}\} \nonumber \\
\mathcal{B}:= & \{[\tilde{\pi}^{\perp\perp}]\cosh\frac{[\alpha]}{2}
-2\bar{l}\sinh\frac{[\alpha]}{2}\}
\label{AB}
\end{align}
\end{equation}
The formal solution for $\bar{\alpha}$ is 
\begin{equation}
\bar{\alpha}=-\tanh^{-1}\frac{\mathcal{B}}{\mathcal{A}}
\label{soluzalfa}
\end{equation}
(we postpone every discussion on the conditions of existence). 
We can now translate this equation (expressed in auxiliary variables) 
into a phase space variables relation. Because of the structure of 
$\mathcal{A}$ and $\mathcal{B}$ (eq. \ref{AB}), the relation (\ref{soluzalfa})
can be written as 
\begin{equation}
\alpha_{+}+\alpha_{-}=f(\alpha_{+}-\alpha_{-})
\end{equation}
and this means
\begin{equation}
\alpha_{+}=F(\alpha_{-})
\end{equation}
where $F$ is an opportune differentiable function whose explicit structure 
is not important at the moment. Keeping present the expression (\ref{ArcTh})
for $\alpha_{\pm}$, we can write
\begin{equation}
\tanh^{-1}\frac{N_{+}^{\perp}}{N_{+}}=F\left(\tanh^{-1}\frac{N_{-}^{\perp}}{N_{-}}\right)
\end{equation}
and then
\begin{equation}
\frac{N_{+}^{\perp}}{N_{+}}=\tanh F\left(\tanh^{-1}\frac{N_{-}^{\perp}}{N_{-}}\right)
=:G\left(\frac{N_{-}^{\perp}}{N_{-}}\right).
\end{equation}
This is the "one more" relation between $N_{\pm}^{\perp}$, $N_{\pm}$, cited in the 
introduction of this chapter, dictated by the solution of the $\alpha$-C.
We will use these relations to write down, in the next section, the hamiltonian
$\mathcal{H}_{118}$ as a function of phase space variables only.

\section{Substitution of $\bar{\alpha}$ back into 
$\mathcal{H}_{118}(\bar{\alpha}, [\alpha])$}

In this section we insert the solution $\bar{\alpha}=\bar{\alpha}([\alpha])$ back
into the hamiltonian and we see how the "new" hamiltonian, modified in such way, does no
more generate the singular constraint $\alpha$-C.
A bit of elaboration permits to write the expression (\ref{Halfa})
of $\mathcal{H}_{118}(\bar{\alpha}, [\alpha])$ as
\begin{multline}
\mathcal{H}_{118}(\bar{\alpha}, [\alpha]) =
-\frac{1}{8 \pi G} \int_{S\cap\Sigma}d^{2}\eta\sqrt{\lambda}
\{\nu(\mathcal{A}\cosh\bar{\alpha}+\mathcal{B}\sinh\bar{\alpha}) \\
+\nu^{K}
((\tilde{\pi}_{K}^{\perp +}+(\bar{\alpha}+\frac{1}{2}[\alpha])_{,K})
-(\tilde{\pi}_{K}^{\perp -}
+(\bar{\alpha}-\frac{1}{2}[\alpha])_{,K}))\}. 
\end{multline}  
where $\mathcal{A}$ and $\mathcal{B}$ have been defined in the previous section.
Reminding that the solution of the $\alpha$-C is 
$\tanh\bar{\alpha}=-(\mathcal{B}/\mathcal{A})$ and 
using a bit of hyperbolic trigonometry we can evaluate the first piece of 
$\mathcal{H}_{118}$. 
We find
\begin{equation}
 \mathcal{A}\cosh\bar{\alpha}+\mathcal{B}\sinh\bar{\alpha}= 
\sqrt{\mathcal{A}^{2}-\mathcal{B}^{2}}.
\end{equation}  
And then the $\alpha$-sector of the "new" hamiltonian is
\begin{equation}
\begin{align}
\mathcal{H}_{118}(\bar{\alpha}, [\alpha]) =
& -\frac{1}{8 \pi G} \int_{S\cap\Sigma}d^{2}\eta\sqrt{\lambda}
\{\nu \sqrt{\mathcal{A}^{2}-\mathcal{B}^{2}}
 & +\nu^{K}
([\tilde{\pi}_{K}^{\perp}]+[\alpha]_{,K})\}. 
\end{align}
\end{equation} 
As we see, $\bar{\alpha}$ has now completely disappeared from this hamiltonian, 
therefore $\delta\mathcal{H}_{118}/\delta\bar{\alpha}=0$. In other words, 
the "bad", singular constraint $\alpha$-C is automatically satisfied and 
it does not appear anymore. It has been radically eliminated from the final 
equations.\\
Finally, the new hamiltonian $\mathcal{H}_{118}$ ($\alpha$-sector) 
can be also written, more properly, 
as a functional of the free phase space variables $N_{-}^{\perp}$, $N_{-}$.
We observe infact that
\begin{multline}
[\alpha]=\alpha_{+}-\alpha_{-}=F(\alpha_{-})-\alpha_{-}= \\ 
=F\left(\tanh^{-1}\frac{N_{-}^{\perp}}{N_{-}}\right)-
\tanh^{-1}\frac{N_{-}^{\perp}}{N_{-}}=:E\left(\frac{N_{-}^{\perp}}{N_{-}}\right).
\end{multline}
Hence
\begin{equation}
\begin{align}
\mathcal{A}=\mathcal{A}([\alpha])=
\mathcal{A}(E\left(\frac{N_{-}^{\perp}}{N_{-}}\right)) \nonumber \\
\mathcal{B}=\mathcal{B}([\alpha])=
\mathcal{B}(E\left(\frac{N_{-}^{\perp}}{N_{-}}\right)).
\end{align}
\end{equation}
Therefore
\begin{equation}
\begin{align}
\mathcal{H}_{118}(N_{-}^{\perp},N_{-}) =
& -\frac{1}{8 \pi G} \int_{S\cap\Sigma}d^{2}\eta\sqrt{\lambda}
\{\sqrt{(N_{-})^{2}-(N_{-}^{\perp})^{2}} \sqrt{\mathcal{A}^{2}-\mathcal{B}^{2}} \nonumber \\
& +\lambda^{KL}N_{L}
([\tilde{\pi}_{K}^{\perp}]+ E\left(\frac{N_{-}^{\perp}}{N_{-}}\right)_{,K})\}. 
\label{newHamN-}
\end{align}
\end{equation} 

\section{A general theorem}
\label{GT}
We could now ask ourselves if the substitution in $\mathcal{H}_{118}$ of  
$\bar{\alpha}=\bar{\alpha}([\alpha])$ preserve correctly the other equations
of motion and constraints. To this question gives an answer the following
\\ \\
\textbf{Theorem}$\;$\textit{The substitution back into the hamiltonian of the 
solution of a constraint 
does not affect the other equations of motion or constraints. In other words, 
we still obtain from the "new" hamiltonian the equations corresponding 
to the old ones.} \\ \\
\textbf{Proof}:$\;$ We conduct the proof using a notation borrowed from our 
special case, but the proof itself is completely general.\\ \\
a) We start writing down in a compact notation the hamiltonian    
$\mathcal{H}_{118}(\bar{\alpha}, [\alpha])$, the variation of $\mathcal{H}_{118}$
in respect to $\bar{\alpha}$ and $[\alpha]$, and comparing it with the symplectic
form $\delta\mathcal{H}_{107}$.

\begin{equation}
\begin{align}  
\delta\mathcal{H}_{118}(\bar{\alpha}, [\alpha]) & =
\frac{\delta\mathcal{H}_{118}}{\delta\bar{\alpha}}\delta\bar{\alpha}+
\frac{\delta\mathcal{H}_{118}}{\delta[\alpha]}\delta[\alpha]
\stackrel{must be}{=}\delta\mathcal{H}_{107} \nonumber \\
& =0\cdot\delta\bar{\alpha}+A\delta[\alpha]
\end{align}
\end{equation}
b) The constraint (129HK), $\alpha$-C, and the equation of motion (131HK) 
are therefore
\begin{equation}
\begin{align}  
& \frac{\delta\mathcal{H}_{118}}{\delta\bar{\alpha}}=C_{129}(\bar{\alpha}, [\alpha])
=0 \nonumber \\
& \frac{\delta\mathcal{H}_{118}}{\delta[\alpha]}=E_{131}(\bar{\alpha}, [\alpha])=A
\end{align}
\end{equation}
c) Now, solve for $\bar{\alpha}$ the constraint $C_{129}$: 
$\bar{\alpha}=\bar{\alpha}([\alpha])$ and substitute the solution back into the 
hamiltonian 
\begin{equation}
\tilde{\mathcal{H}}_{118}:=\mathcal{H}_{118}(\bar{\alpha}([\alpha]), [\alpha])
\nonumber
\end{equation}
d) Calculate from this "new" hamiltonian $\tilde{\mathcal{H}}_{118}$ the "new" 
equations of motion and constraints

\begin{equation}
\begin{align}  
& \frac{\delta\tilde{\mathcal{H}}_{118}}{\delta\bar{\alpha}}\equiv 0
\stackrel{must be}{=}0 \nonumber \\
& \frac{\delta\tilde{\mathcal{H}}_{118}}{\delta[\alpha]}=
\frac{\delta\mathcal{H}_{118}}{\delta\bar{\alpha}}
\frac{\delta\bar{\alpha}}{\delta[\alpha]}+
\frac{\delta\mathcal{H}_{118}}{\delta[\alpha]}\stackrel{must\;be}{=}A
\label{neweq}
\end{align}
\end{equation}
So, the first equation says that the $\alpha$-C is automatically satisfied 
(because $\tilde{\mathcal{H}}_{118}$ doesn't depend on $\bar{\alpha}$).
For the second one, observe that now, after the substitution, we have

\begin{equation}
\frac{\delta\mathcal{H}_{118}}{\delta\bar{\alpha}}=
C_{129}(\bar{\alpha}([\alpha]), [\alpha])\equiv 0
\end{equation}
where the last equation is an identity because $\bar{\alpha}([\alpha])$
\textit{is} the solution of the constraint $C_{129}$.
Besides now
\begin{equation}
\frac{\delta\mathcal{H}_{118}}{\delta[\alpha]}=
E_{131}(\bar{\alpha}([\alpha]), [\alpha])
\end{equation}
Hence the "new" equations (\ref{neweq}) become

\begin{equation}
\begin{align}  
& \frac{\delta\tilde{\mathcal{H}}_{118}}{\delta\bar{\alpha}}\equiv 0
=0 \nonumber \\
& \frac{\delta\tilde{\mathcal{H}}_{118}}{\delta[\alpha]}
\equiv E_{131}(\bar{\alpha}([\alpha]), [\alpha])=A
\end{align}
\end{equation}
e) This means that:\\ \\ 
i) the first equation, the constraint, is automatically 
satisfied;\\ \\
ii) the second equation (eq. 131HK in our particular case) becomes
\begin{equation}
E_{131}(\bar{\alpha}([\alpha]), [\alpha])=A
\end{equation}
which is the \textit{the same equation} as in point b), just only with 
$\bar{\alpha}$ replaced with $\bar{\alpha}([\alpha])$.
Thus, we conclude that the elimination of the constraint does not affect 
the other equations.

\section{Boundary conditions for lapse $N$ and shift $N^{\perp}$ functions}

The relations $N^{\perp}_{\pm}=\nu\sinh\alpha_{\pm}$, $\;$
$N_{\pm}=\nu\cosh\alpha_{\pm}$ (dictated to us by the choice of adapted coordinates) 
can be interpretated as a sort of boundary conditions for $N$ and $N^{\perp}$.
The fields $N(x)$, $N^{\perp}(x)$ can fluctuate freely (following their equations 
of motion, of course) when $x\in S^{\pm}$, but they must assume the values 
$\nu\cosh\alpha_{\pm}$, $\nu\sinh\alpha_{\pm}$ (respectively) when $x\longrightarrow
\Sigma\cap S^{\pm}$. These conditions at the boundary $\Sigma$ are analogous to 
fall-off conditions at infinity, or control conditions. We have seen that their role 
is to define the configuration space of our system just as control modes or 
fall-off conditions do.\\
We have seen that the phase space variables 
$N_{\pm}$, $N_{\pm}^{\perp}$ are not free, but they are linked by the relation
\begin{equation}
(N_{+})^{2}-(N_{+}^{\perp})^{2}=(N_{-})^{2}-(N_{-}^{\perp})^{2}.
\end{equation}
From the solution 
of the $\alpha$-C we obtained a further relation
\begin{equation}
\frac{N_{+}^{\perp}}{N_{+}}=G\left(\frac{N_{-}^{\perp}}{N_{-}}\right).
\end{equation}
Therefore the relevant phase space variables really free are
\begin{equation}
N_{-}, \; N_{-}^{\perp}, \; N_{1}, \; N_{2} \nonumber
\end{equation}     
The new hamiltonian (\ref{newHamN-}) has been written as a functional of these free
phase space variables.\\
We note that also $N_{+}^{\perp}$ and $N_{+}$ can be written as functions 
of $N_{-}, \; N_{-}^{\perp}$. In fact we have
\[ \left\{ \begin{array}{ll}
(N_{+})^{2}-(N_{+}^{\perp})^{2}=(N_{-})^{2}-(N_{-}^{\perp})^{2} \\ \\
\frac{N_{+}^{\perp}}{N_{+}}=G\left(\frac{N_{-}^{\perp}}{N_{-}}\right)
\end{array}
\right. \] 
and from here
\begin{equation}
\begin{align}
N_{+} & =\left(\frac{(N_{-})^{2}-(N_{-}^{\perp})^{2}}
{1-G(N_{-}^{\perp}/N_{-})^{2}}\right)^{1/2}=:N_{+}(N_{-},\;N_{-}^{\perp}) \nonumber \\
N_{+}^{\perp} & = \left(\frac{(N_{-})^{2}-(N_{-}^{\perp})^{2}}
{1-G(N_{-}^{\perp}/N_{-})^{2}}\right)^{1/2}\cdot 
G\left(\frac{N_{-}^{\perp}}{N_{-}}\right)=:N_{+}^{\perp}(N_{-},\;N_{-}^{\perp}) 
\label{N+N-}
\end{align}
\end{equation}   
Expressed in these variables, of course the $\alpha$-C is now an identity
\begin{equation}
[N\tilde{\pi}^{\perp \perp}-lN^{\perp}] =
\tilde{\pi}^{\perp \perp}_{+}N_{+}(N_{-},\;N_{-}^{\perp})-
l_{+}N_{+}^{\perp}(N_{-},\;N_{-}^{\perp})
 -\tilde{\pi}^{\perp \perp}_{-}N_{-}+
l_{-}N_{-}^{\perp}\equiv 0.
\end{equation}  
Using these variables, \underline{the new hamiltonian} can be written with an 
$\alpha$-sector form more agile than in (\ref{newHamN-}). It is
\begin{equation}
\begin{align} 
\mathcal{H}_{118}^{NEW(2)} 
& =\frac{1}{16 \pi G} \int_{S^{-}\cup S^{+}}d^{3}y\{NC+N^{k}C_{k}\} \nonumber \\
&-\frac{1}{8 \pi G} \int_{S\cap\Sigma}d^{2}\eta\sqrt{\lambda}
\{[\tilde{\pi}^{\perp\perp}N^{\perp}-lN] \nonumber \\
& +\lambda^{KL}N_{L}
([\tilde{\pi}_{K}^{\perp}]+ E\left(\frac{N_{-}^{\perp}}{N_{-}}\right)_{,K})\}
-\int_{S\cap\Sigma}d^{2}\eta T^{0}_{s\;0}
\label{H118N2}
\end{align}
\end{equation}
where, of course, now it is understood that
$N_{+}(N_{-},\;N_{-}^{\perp})$, $N_{+}^{\perp}(N_{-},\;N_{-}^{\perp})$
and these are the functions given before in (\ref{N+N-}). \\    
\underline{The phase space} is now described by the variables \\ \\
\( \begin{array}{cll}
\hspace*{3cm}-\frac{1}{16 \pi G}\;q_{kl} & \longleftrightarrow & \pi^{kl} \\ \\
\hspace*{3cm}-\frac{1}{8 \pi G}\sqrt{\lambda} & \longleftrightarrow & 
E\left(\frac{N_{-}^{\perp}}{N_{-}}\right) \\ \\
\hspace*{3cm}z^{A} & \longleftrightarrow & p_{A}
\end{array} \) \\ \\  
\underline{The new symplectic form} is
\begin{equation}
\begin{align}
\delta\mathcal{H}^{SYM(2)}_{107(NEW)} & = \frac{1}{16 \pi G}\int_{S}d^{3}y(-\dot{\pi}^{kl}
\delta q_{kl}+\dot{q}_{kl}\delta\pi^{kl}) \nonumber \\
& +\frac{1}{16 \pi G}\int_{S\cap\Sigma}d^{2}
\eta\sqrt{\lambda}\lambda^{KL}  (-\dot{E}\left(\frac{N_{-}^{\perp}}{N_{-}}\right)\delta\lambda_{KL}
+\dot{\lambda}_{KL}\delta E\left(\frac{N_{-}^{\perp}}{N_{-}}\right)) \nonumber \\ 
& -\int_{S\cap\Sigma}d^{2}\eta(\dot{p}_{A}\delta
z^{A}-\dot{z}^{A}\delta p_{A}).
\end{align}
\end{equation}
From here we read the new phase space variables that are those just showed.\\
\underline{New variation of $\mathcal{H}_{118}$}:
The calculus of the variation of the last version of $\mathcal{H}_{118}$ proceeds
by keeping in mind the variation (\ref{varH118new}) and the usual identities 
$\lambda_{KL}=q_{KL}|_{S\cap\Sigma}$, $\nu^{K}=\lambda^{KL}N_{L}$, 
$[\alpha]=[\tanh^{-1}(N^{\perp}/N)]=E(N^{\perp}_{-}/N_{-})$, 
$\nu=((N_{-})^{2}-(N_{-}^{\perp})^{2})^{1/2}$. Note that now $N_{+}$, 
$N_{+}^{\perp}$ have been expressed by $N_{-}$, $N_{-}^{\perp}$. 
The variation is
\begin{equation}
\begin{align}
\delta\mathcal{H}_{118}^{NEW(2)} & =\frac{1}{16 \pi G}\int_{S}d^{3}y(C_{k}\delta N^{k}+
C\delta N + a^{kl}\delta q_{kl} + b_{kl}\delta\pi^{kl}) \nonumber \\
& +\frac{1}{16 \pi G}
\int_{S\cap\Sigma}d^{2}\eta[2REST-\sqrt{\lambda}B^{k}m_{k}] \nonumber \\
& -\frac{1}{8 \pi G}\int_{S\cap\Sigma}d^{2}\eta\frac{\sqrt{\lambda}}
{\sqrt{|\gamma|}}
\{\left([Q^{\perp\perp}]-8\pi G T^{\perp\perp}_{s}\right)
\delta\left(\sqrt{(N_{-})^{2}-(N_{-}^{\perp})^{2}}\right) \nonumber \\
& +\left([Q^{\perp}_{K}]-8 \pi GT^{\perp}_{s\;K}\right)\delta(\lambda^{KL}N_{L})\}
- \int_{S\cap\Sigma}d^{2}\eta\left\{\frac{1}{2}T^{KL}_{s}\delta\lambda_{KL}\right\} \nonumber \\
& - \int_{S\cap\Sigma}d^{2}\eta\left\{\left(\frac{\partial T^{0}_{s\;0}}
{\partial z^{A}}-\frac{\partial}{\partial\eta^{M}}\frac{\partial T^{0}_{s\;0}}
{\partial z^{A}_{M}}\right)\delta z^{A}+ \frac{\partial T^{0}_{s\;0}}{\partial 
p_{A}}\delta p_{A}\right\}.
\label{varH118N2}
\end{align}
\end{equation} \\
\underline{Old and new equations}: The comparison between
\begin{equation}
\delta\mathcal{H}_{118}^{NEW(2)}=\delta\mathcal{H}^{SYM(2)}_{107(NEW)}
\label{SCE}
\end{equation}   
gives us the standard canonical equations

\[ \left\{ \begin{array}{ll}
C_{k}=0 \\ \\
C=0
\end{array}
\right. \quad \left\{ \begin{array}{ll}
\dot{q}_{kl}=b_{kl} \\ \\
\dot{\pi}^{kl}=-a^{kl}
\end{array}
\right.  \quad \left\{ \begin{array}{ll} 
[Q^{\perp\perp}]=8 \pi G T^{\perp\perp}_{s} \\ \\

[Q^{\perp}_{K}]= 8 \pi G T^{\perp}_{s\;K}
\end{array}
\right.    \quad \left\{ \begin{array}{ll}
\dot{p}_{A}=\frac{\partial T^{0}_{s\;0}}
{\partial z^{A}}-\frac{\partial}{\partial\eta^{M}}\frac{\partial T^{0}_{s\;0}}
{\partial z^{A}_{M}} \\ \\
-\dot{z}^{A}=\frac{\partial T^{0}_{s\;0}}{\partial p_{A}}
\end{array}
\right.    \] \\ \\
still unchanged in form.\\
The remaining "REST" terms to be compared are
\begin{equation}
\begin{align}
& \int_{S\cap\Sigma}d^{2}\eta[2REST-\sqrt{\lambda}B^{k}m_{k}]
-\int_{S\cap\Sigma}d^{2}\eta \{8 \pi G T^{KL}_{s}\delta\lambda_{KL}\} = \nonumber \\
& \int_{S\cap\Sigma}d^{2}
\eta\sqrt{\lambda}\lambda^{KL}(-\dot{E}\left(\frac{N_{-}^{\perp}}{N_{-}}\right)\delta\lambda_{KL}
+\dot{\lambda}_{KL}\delta E\left(\frac{N_{-}^{\perp}}{N_{-}}\right))   
\end{align}
\end{equation}
which explicitly read (see variation (\ref{var1.42}))
\begin{equation}
\begin{align}
& \int_{S\cap\Sigma}d^{2}\eta\{-\sqrt{\lambda}([\tilde{\pi}^{KL}
N^{\perp}+l^{KL}N-l\lambda^{KL}N-N_{,k}m^{k}\lambda^{KL}] \nonumber \\
& +\lambda^{MR}N_{R}\lambda^{KL}E(\frac{N_{-}^{\perp}}{N_{-}})_{,M})\delta\lambda_{KL}
-2\sqrt{\lambda}[\tilde{\pi}^{\perp\perp}N-lN^{\perp}]
\delta(\tanh^{-1}\frac{N^{\perp}_{\pm}}{N_{\pm}}) \nonumber \\
& +2\sqrt{\lambda}(-\tilde{\pi}^{\perp\perp}N+lN^{\perp}
+(\lambda^{KL}N_{L})_{||K})^{\pm}\delta E(\frac{N_{-}^{\perp}}{N_{-}})\}
-\int_{S\cap\Sigma}d^{2}\eta(8 \pi G T^{KL}_{s}\delta\lambda_{KL}) \nonumber \\
& =\int_{S\cap\Sigma}d^{2}\eta
\sqrt{\lambda}\lambda^{KL}\{-\dot{E}\left(\frac{N_{-}^{\perp}}{N_{-}}\right)\delta\lambda_{KL}
+\dot{\lambda}_{KL}\delta E\left(\frac{N_{-}^{\perp}}{N_{-}}\right)\}
\label{varH118rest}
\end{align}
\end{equation}  
We are varing $\delta\mathcal{H}_{118}^{NEW(2)}$, therefore we are using the variables \\
$N_{+}=N_{+}(N_{-},\;N_{-}^{\perp})$, 
$N_{+}^{\perp}=N_{+}^{\perp}(N_{-},\;N_{-}^{\perp})$,
and these variables satisfy, by construction, the identity
\begin{equation}
[\tilde{\pi}^{\perp\perp}N-lN^{\perp}]=0.
\end{equation}
Hence the second line of the last equation semplifies and finally,
from the comparison, we get the equations: \\ \\
1)- The new form of the equation of motion for $\lambda$ (131HK)
\begin{equation}
\dot{\lambda}=2 \lambda(-\tilde{\pi}^{\perp\perp}N+lN^{\perp}
+(\lambda^{KL}N_{L})_{||K})^{\pm}
\end{equation}
where now it is understood that $N_{+}=N_{+}(N_{-},\;N_{-}^{\perp})$, 
$N_{+}^{\perp}=N_{+}^{\perp}(N_{-},\;N_{-}^{\perp})$ \\ \\
2)- The new form of the equation 130HK
\begin{equation}
\begin{align}
& -\sqrt{\lambda}([\tilde{\pi}^{KL}
N^{\perp}+l^{KL}N-l\lambda^{KL}N-N_{,k}m^{k}\lambda^{KL}]
+\lambda^{MR}N_{R}\lambda^{KL}E(\frac{N_{-}^{\perp}}{N_{-}})_{,M}) \nonumber \\
& -8 \pi G T^{KL}_{s}= 
-\sqrt{\lambda}\lambda^{KL}\dot{E}\left(\frac{N_{-}^{\perp}}{N_{-}}\right) 
\label{new130HK2}
\end{align}
\end{equation}
also here of course it is understood that $N_{+}$, $N_{+}^{\perp}$ are given by 
(\ref{N+N-}).\\
The trace of the last equation is the equation of motion for the variable $E(N_{-}^{\perp}/N_{-})$
\begin{equation}
\begin{align}
\sqrt{\lambda}\dot{E}\left(\frac{N_{-}^{\perp}}{N_{-}}\right)
& =4 \pi G T^{KL}\lambda_{KL}
+\frac{1}{2}\sqrt{\lambda}([\tilde{\pi}^{KL}\lambda_{KL}N^{\perp}-lN-2N_{,k}m^{k}] \nonumber \\
& +2\lambda^{MR}N_{R}E(\frac{N_{-}^{\perp}}{N_{-}})_{,M}) 
\label{new137HK}
\end{align}
\end{equation} 
while the trace free part of the same equation is the "new" version of the $\lambda$-C
\begin{equation}
\begin{align}
& -\sqrt{\lambda}[N^{\perp}(\tilde{\pi}^{KL}-\frac{1}{2}\tilde{\pi}^{MN}
\lambda_{MN}\lambda^{KL})+N(l^{KL}-\frac{1}{2}l\lambda^{KL})]= \nonumber \\
& = 8 \pi G (T^{KL}-\frac{1}{2}T^{MN}\lambda_{MN}\lambda^{KL})
\end{align}
\end{equation}
where now it is understood, as usual, that $N_{+}$, $N_{+}^{\perp}$ are given by 
the equations (\ref{N+N-}) and, of course, $\lambda_{KL}=q_{KL}|_{\Sigma}$.
Note that the $\lambda$-C is still formally unchanged even after the last 
substitutions. 
The only constraint that now remains to be solved is the $\lambda$-C: 
it will be dealt with in the next chapter.    
%
%
%
%
%
%
\chapter{Explicit solution of the $\lambda$ Constraint}
In this chapter we will deal with the explicit solution of the $\lambda$-C. 
This is in fact a singular constraint and in order to get rid of it we will adopt
a strategy analogous to that adopted for the $\alpha$-C. First of all, we will 
show that the equation (\ref{new130HK2})(i.e. the new form of the equation
130HK) can be obtained by varying $\mathcal{H}_{118}^{NEW(2)}$ 
in respect to $\lambda_{KL}$ (here and after, $\lambda_{KL}=q_{KL}|_{\Sigma}$) 
and comparing this variation with the relevant part 
of the symplectic form. The equation 130HK is important because its tracefree
part \textit{is} the $\lambda$-C.\\
Then, we will propose an opportune splitting in two parts of the variable 
$\lambda_{KL}$, in such a way that the variation of $\mathcal{H}_{118}^{NEW(2)}$ in 
respect to the second part of $\lambda_{KL}$ give directly the $\lambda$-C.
This procedure is analogous to that used for the $\alpha$-C, when we showed
that the $\alpha$-C can be obtained directly by varying the hamiltonian in 
respect to the variable $\bar{\alpha}$.
Finally we will solve explicitly the $\lambda$-C in respect to that variable 
which permits (by variation of $\mathcal{H}_{118}^{NEW(2)}$) to obtain the 
$\lambda$-C itself. 
Moreover we'll substitute this explicit solution in the shell part of
$\mathcal{H}_{118}^{NEW(2)}$  and we'll discuss the boundary conditions 
for $q_{AB}$ on the $\Sigma$ surface.

\section{Variation of $\mathcal{H}_{118}^{NEW(2)}$ in respect to $\lambda_{KL}$}

We should remind here the expression (\ref{H118N2}) of $\mathcal{H}_{118}^{NEW(2)}$  
and with the help of the total variation (eqs. \ref{varH118N2}, \ref{varH118rest})
we can write down the 
variation of $\mathcal{H}_{118}^{NEW(2)}$ in respect to $\lambda_{KL}$. 
Using only the pieces of $\delta\mathcal{H}_{118}^{NEW(2)}$ containing 
$\delta\lambda_{KL}$ (to choose them, have a look, for example, 
to (\ref{varH118rest})), we can write
\begin{equation}
\begin{align}
\delta\mathcal{H}_{118}^{NEW(2)} & (/\delta\lambda_{KL})=
\frac{1}{16 \pi G} \int_{S\cap\Sigma}d^{2}\eta
\{-\sqrt{\lambda}([\tilde{\pi}^{KL}N^{\perp}
+l^{KL}N-l\lambda^{KL}N \nonumber \\ \nonumber \\          
& -N_{,k}m^{k}\lambda^{KL}] 
+\lambda^{KL}N^{M}E\left(\frac{N_{-}^{\perp}}{N_{-}}\right)_{,M})
-8 \pi G T^{KL}\}\delta\lambda_{KL}
\label{varHlambda1}
\end{align}
\end{equation}
This must be compared with the relevant part (i.e. in $\delta\lambda_{KL}$)
of the symplectic form
\begin{equation}
\delta\mathcal{H}_{107(NEW)}^{SYM(2)}
=-\frac{1}{16 \pi G} \int_{S\cap\Sigma}d^{2}\eta\sqrt{\lambda}
\dot{E}\left(\frac{N_{-}^{\perp}}{N_{-}}\right)
\lambda^{KL}\delta\lambda_{KL}
\label{varHlambda2}
\end{equation}
Therefore we get the equation
\begin{equation}
\begin{align}
& -\sqrt{\lambda}
([\tilde{\pi}^{KL}N^{\perp}+l^{KL}N-l\lambda^{KL}N-N_{,k}m^{k}\lambda^{KL}]
+\lambda^{KL}N^{M}E\left(\frac{N_{-}^{\perp}}{N_{-}}\right)_{,M})
\nonumber \\
& -8 \pi G T^{KL}
=-\sqrt{\lambda}\lambda^{KL}\dot{E}\left(\frac{N_{-}^{\perp}}{N_{-}}\right)
\label{130HK}
\end{align}
\end{equation} 
and this is the new form of equation 130HK already stated in (\ref{new130HK2})
(of course here $N_{+}$, $N_{+}^{\perp}$ are given by (\ref{N+N-})).
So, we have proved that equation 130HK comes from the variation of 
$\mathcal{H}_{118}^{NEW(2)}$ in respect to $\lambda_{KL}$. 
The trace-free part of 130HK
is the so called $\lambda$-C, while the trace of 130HK is the equation of 
motion 137HK (whose new form is stated in (\ref{new137HK})).\\ 
If we define
\begin{equation}
\begin{align}
H^{KL}:= & -\sqrt{\lambda}([\tilde{\pi}^{KL}N^{\perp}+l^{KL}N-l\lambda^{KL}N-N_{,k}m^{k}\lambda^{KL}] \nonumber \\
& +\lambda^{KL}\lambda^{MR}N_{R}E\left(\frac{N_{-}^{\perp}}{N_{-}}\right)_{,M})-8 \pi G T^{KL} \nonumber \\
A^{KL}:= & -\sqrt{\lambda}\lambda^{KL}\dot{E}\left(\frac{N_{-}^{\perp}}{N_{-}}\right)
\end{align}
\end{equation}
the equation (\ref{varHlambda1})=(\ref{varHlambda2}) can be written as 
\begin{equation}
\begin{align}
\delta\mathcal{H}_{118}^{NEW(2)}(/\delta\lambda_{KL}) & =\frac{1}{16 \pi G}
\int_{S\cap\Sigma}d^{2}\eta H^{KL}\delta\lambda_{KL} \nonumber \\
& =\frac{1}{16 \pi G} \int_{S\cap\Sigma}d^{2}\eta A^{KL}\delta\lambda_{KL}=
\delta\mathcal{H}_{107(NEW)}^{SYM(2)}
\end{align}
\end{equation}
which means
\begin{equation}
\int_{S\cap\Sigma}d^{2}\eta (H^{KL}-A^{KL})\delta \lambda_{KL}=0.
\label{*}
\end{equation}
Because of the arbitrariness of the variations $\delta\lambda_{KL}$, the 
last equation is equivalent to
\begin{equation}
H^{KL}=A^{KL}
\end{equation}
which is the equation (\ref{130HK}).

\section{Splitting of $\lambda_{KL}$}

We propose now a splitting of $\lambda_{KL}$ in two different parts, 
in a way such that 
the constraint $\lambda$-C 
can be obtained with a variation of $\mathcal{H}_{118}^{NEW(2)}$
in respect to the second part of $\lambda_{KL}$.
The splitting proposed is the following  
\begin{equation}
\lambda_{KL}=\sqrt{\lambda}\;\kappa_{KL}
\end{equation}
where $\lambda=\textup{det}\lambda_{KL}$, and $\kappa_{KL}$ is a $2\times2$ 
unimodular matrix (i.e. det$\kappa_{KL}=1$). Hence we have
\begin{equation}
\textup{det}\lambda_{KL}=\textup{det}(\sqrt{\lambda}\;\kappa_{KL})
=\lambda\;\textup{det}\kappa_{KL}=\lambda.
\end{equation}
Moreover $\lambda_{KL}$ is made of $3$ free parameters (it is a symmetric tensor).
So $\kappa_{KL}$ is a symmetric unimodular matrix with $2$ free parameters.\\
Because of the splitting we have
\begin{equation}
\begin{align}
& \lambda_{KL}=\sqrt{\lambda}\;\kappa_{KL}, \quad 
\lambda^{KL}=\frac{1}{\sqrt{\lambda}}\;\kappa^{KL} \nonumber \\
& \kappa=1 \quad \Longrightarrow \quad \kappa^{KL}\delta\kappa_{KL}=0.
\end{align}
\end{equation}
This means
\begin{equation}
\delta \lambda_{KL}=\frac{1}{2\sqrt{\lambda}}\kappa_{KL}\delta\lambda
+\sqrt{\lambda}\delta\kappa_{KL}.
\end{equation}
Equation (\ref{*}) becomes
\begin{equation}
\int_{S\cap\Sigma}d^{2}\eta
\left\{\frac{1}{2\sqrt{\lambda}}(H^{KL}-A^{KL})\kappa_{KL}\delta\lambda
+\sqrt{\lambda}(H^{KL}-A^{KL})\delta\kappa_{KL}\right\}=0.
\label{**}
\end{equation}
Yet, the variables $\kappa_{KL}$ are not independent, because 
$\kappa=1$ (or, in differential version, $\kappa^{KL}\delta\kappa_{KL}=0$).
Therefore we cannot infer, at this stage, a finite equation from (\ref{**}).
In order to get this result, let's parametrize $\kappa_{KL}$. We write down
a $C^{\infty}$ representation of this unimodular matrix, by
writing the elements of $\kappa_{KL}$ as functions of two free parameters.
The matrix $\kappa_{KL}$ is symmetric and unimodular, 
$\kappa_{11}\kappa_{22}-(\kappa_{12})^{2}=1$. \\
The form of the last relation suggests to use functions as $\cosh$, $\sinh$.
An accurate inspection brings us to the $C^{\infty}$ representation
\[ \kappa_{KL} =\left[
\begin{array}{ll}
e^{\phi}\cosh\chi & \sinh\chi  \\    
\sinh \chi & e^{-\phi}\cosh\chi 
\end{array} 
\right]   \]
Of course this representation satisfies automatically the constraint $\kappa=1$
and its differential version $\kappa^{KL}\delta\kappa_{KL}=0$.
Because $\kappa_{KL}=\kappa_{KL}(\phi, \chi)$ we have
\begin{equation}
\delta\kappa_{KL}=\frac{\partial\kappa_{KL}}{\partial\phi}\delta\phi+
\frac{\partial\kappa_{KL}}{\partial\chi}\delta\chi
\end{equation}
where $\phi$ and $\chi$ are now free parameters. The equation (\ref{**}) becomes
\begin{equation}
\begin{align}
\int_{S\cap\Sigma}d^{2}\eta
& \{\frac{1}{2\sqrt{\lambda}}(H^{KL}-A^{KL})\kappa_{KL}\delta\lambda
+\sqrt{\lambda}(H^{KL}-A^{KL})\frac{\partial\kappa_{KL}}{\partial\phi}\delta\phi \nonumber \\
& +\sqrt{\lambda}(H^{KL}-A^{KL})\frac{\partial\kappa_{KL}}{\partial\chi}\delta\chi\}=0.
\end{align}
\end{equation}
$\lambda$ , $\phi$ , $\chi$ are now free variables, therefore the last equation is equivalent to the 
three finite equations
\begin{equation}
\begin{align}
\frac{1}{2\sqrt{\lambda}}(H^{KL}-A^{KL})\kappa_{KL} & =0 \nonumber \\
\sqrt{\lambda}(H^{KL}-A^{KL})\frac{\partial\kappa_{KL}}{\partial\phi} & =0 \nonumber \\
\sqrt{\lambda}(H^{KL}-A^{KL})\frac{\partial\kappa_{KL}}{\partial\chi} & =0
\label{3eqs}
\end{align}
\end{equation}
We will prove now that the first equation of (\ref{3eqs}) is equivalent to the new form of the 
equation 137HK (eq. \ref{new137HK}, equation of motion for $E(N_{-}^{\perp}/N_{-})$), 
while the second and the third 
equation are equivalent, together, to the $\lambda$-C.\\

$\bullet$ As regard the first equation, it is sufficient to observe that
\begin{equation}
\lambda^{KL}\cdot\frac{\kappa_{KL}}{2\sqrt{\lambda}}=\frac{1}{\lambda}
\end{equation}
to arrive, with easy calculations, to the equation
\begin{equation}
\begin{align}
\sqrt{\lambda}\dot{E}\left(\frac{N_{-}^{\perp}}{N_{-}}\right) & =4 \pi G T^{KL}\lambda_{KL}
+\frac{1}{2}\sqrt{\lambda}([\tilde{\pi}^{KL}\lambda_{KL}N^{\perp}-lN-2N_{,k}m^{k}] \nonumber \\ 
& +2\lambda^{MR}N_{R}E(\frac{N_{-}^{\perp}}{N_{-}})_{,M}) 
\end{align}
\end{equation} 
which is the new form of the 137HK (eq. \ref{new137HK}).\\

$\bullet$ As regard the second and the third equation, observe, first of all, that
    
\[ \frac{\partial\kappa_{KL}}{\partial\phi} =\left[
\begin{array}{cc}
e^{\phi}\cosh\chi & 0  \\    
0 & -e^{-\phi}\cosh\chi 
\end{array} 
\right]   \]        \\ 
\[ \frac{\partial\kappa_{KL}}{\partial\chi} =\left[
\begin{array}{cc}
e^{\phi}\sinh\chi & \cosh\chi  \\    
\cosh\chi & e^{-\phi}\sinh\chi 
\end{array} 
\right]   \]       \\
\[ \kappa^{KL} =\left[
\begin{array}{cc}
e^{-\phi}\cosh\chi & -\sinh\chi  \\    
-\sinh \chi & e^{\phi}\cosh\chi 
\end{array} 
\right]   \]
and therefore,
\begin{equation}
\begin{align}
& \kappa^{KL}\frac{\partial\kappa_{KL}}{\partial\phi}=0 \nonumber \\ 
& \kappa^{KL}\frac{\partial\kappa_{KL}}{\partial\chi}=0
\end{align}
\end{equation}
This means also
\begin{equation}
\lambda^{KL}\frac{\partial\kappa_{KL}}{\partial\phi}=
\lambda^{KL}\frac{\partial\kappa_{KL}}{\partial\chi}=0
\end{equation}
Hence, the second and the third equation of (\ref{3eqs}) become
\begin{equation}
\begin{align}
\mathcal{A}^{KL}\frac{\partial\kappa_{KL}}{\partial\phi}=0 \nonumber \\
\mathcal{A}^{KL}\frac{\partial\kappa_{KL}}{\partial\chi}=0 
\end{align}
\end{equation}
where
\begin{equation}
\mathcal{A}^{KL}=\sqrt{\lambda}[\tilde{\pi}^{KL}N^{\perp}+Nl^{KL}]+8 \pi G T^{KL}
\end{equation}

Explicitly, the previous conditions read
\begin{equation}
\begin{align}
& \mathcal{A}^{11}e^{\phi}\cosh\chi-\mathcal{A}^{22}e^{-\phi}\cosh\chi=0 \nonumber \\
& 2\mathcal{A}^{12}\cosh\chi+\mathcal{A}^{11}e^{\phi}\sinh\chi+
  \mathcal{A}^{22}e^{-\phi}\sinh\chi=0
\end{align}
\end{equation} 
which, in terms of $\phi$ and $\chi$ give
\begin{equation}
\begin{align}
e^{\phi} & =\left(\frac{\mathcal{A}^{22}}{\mathcal{A}^{11}}\right)^{1/2}\nonumber \\
\tanh\chi & =-\frac{\mathcal{A}^{12}}{\sqrt{\mathcal{A}^{11}\mathcal{A}^{22}}}.
\label{soluz}
\end{align}
\end{equation}
We show now that these two conditions are equivalent to the $\lambda$-C. 
In fact, note that the $\lambda$-C 
\begin{equation}
\begin{align}
\sqrt{\lambda}[N^{\perp}(\tilde{\pi}^{KL}-\frac{1}{2}\tilde{\pi}^{AB}
\lambda_{AB}\lambda^{KL}) & +N(l^{KL}-\frac{1}{2}l\lambda^{KL})] \nonumber \\
& + 8 \pi G (T^{KL}-\frac{1}{2}T^{AB}\lambda_{AB}\lambda^{KL})=0
\end{align}
\end{equation}
represents two equations (or conditions) (the trace of the LHS expression is zero). 
Moreover, it can be written as
\begin{equation}
(\delta^{A}_{K}\delta^{B}_{L}-\frac{1}{2}\kappa_{KL}\kappa^{AB})
\mathcal{A}^{KL}=0
\end{equation}
where $\mathcal{A}^{KL}$ is previously defined.
But this means
\begin{equation}
2\mathcal{A}^{AB}\kappa_{BC}=\kappa_{KL}\mathcal{A}^{KL}\delta^{B}_{C}
\end{equation}
which is equivalent to the linear system of equations
\begin{equation}
\begin{align}
& \mathcal{A}^{11}\kappa_{12}+\mathcal{A}^{12}\kappa_{22}=0 \nonumber \\
& \mathcal{A}^{11}\kappa_{11}-\mathcal{A}^{22}\kappa_{22}=0.
\end{align}
\end{equation}
In terms of the free parameters $\phi$, $\chi$ the equations read
\begin{equation}
\begin{align}
& \mathcal{A}^{11}\sinh\chi+\mathcal{A}^{12}e^{-\phi}\cosh\chi=0 \nonumber \\
& \mathcal{A}^{11}e^{\phi}\cosh\chi-
  \mathcal{A}^{22}e^{-\phi}\cosh\chi=0
\end{align}
\end{equation} 
and their solutions in terms of $\phi$ and $\chi$ are 
\begin{equation}
\begin{align}
e^{\phi} & =\left(\frac{\mathcal{A}^{22}}{\mathcal{A}^{11}}\right)^{1/2}\nonumber \\
\tanh\chi & =-\frac{\mathcal{A}^{12}}{\sqrt{\mathcal{A}^{11}\mathcal{A}^{22}}}.
\end{align}
\end{equation}
Clearly, they coincide with (\ref{soluz}). (QED!)\\ 
So, using the splitting $\lambda_{KL}=\sqrt{\lambda}\;\kappa_{KL}$ we have proved 
that a variation of the hamiltonian $\mathcal{H}_{118}^{NEW(2)}$ in respect
to $\lambda$ gives the equation of motion for $E(N_{-}^{\perp}/N_{-})$
(eq. \ref{new137HK} i.e. 137HK new form), while a variation 
of $\mathcal{H}_{118}^{NEW(2)}$ in respect to the free parameters 
$\phi$, $\chi$ (on which $\kappa_{KL}$ depends, 
$\kappa_{KL}=\kappa_{KL}(\phi, \chi)$) 
gives the constraint $\lambda$-C.\\
The next step is to solve the $\lambda$-C in respect to the variables $\phi$, $\chi$
(which generate it by variation): indeed, this has been just done!

\section{Explicit solution of the $\lambda$-C}

In this section we summarize the solution of the $\lambda$-C already
calculated in the previous section. We have seen that the $\lambda$-C can be 
expressed as 
\begin{equation}
(\delta^{A}_{K}\delta^{B}_{L}-\frac{1}{2}\kappa_{KL}\kappa^{AB})
\mathcal{A}^{KL}=0
\end{equation}
(where $\kappa_{KL}=\kappa_{KL}(\phi, \chi)$) and the solution in terms of 
the free parameters $\phi$ and $\chi$ is
\begin{equation}
\begin{align}
e^{\phi} & =\left(\frac{\mathcal{A}^{22}}{\mathcal{A}^{11}}\right)^{1/2}\nonumber \\
\tanh\chi & =-\frac{\mathcal{A}^{12}}{\sqrt{\mathcal{A}^{11}\mathcal{A}^{22}}}.
\end{align}
\end{equation}
We can also write down an explicit solution in terms of the matrix elements 
$\kappa_{AB}$. In fact, starting again from (\ref{soluz}) and using hyperbolic trigonometry
we get
\begin{equation}
\begin{align}
\sinh\chi & =-\frac{\mathcal{A}^{12}}{\sqrt{\mathcal{A}}}\nonumber \\
\cosh\chi & =\frac{\sqrt{\mathcal{A}^{11}\mathcal{A}^{22}}}{\sqrt{\mathcal{A}}}
\end{align}
\end{equation}
where $\mathcal{A}=\textup{det}\mathcal{A}^{KL}=
\mathcal{A}^{11}\mathcal{A}^{22}-(\mathcal{A}^{12})^{2}$.
Therefore finally
\begin{equation}
\begin{align}
\kappa_{11} & =\frac{\mathcal{A}^{22}}{\sqrt{\mathcal{A}}}\nonumber \\
\kappa_{22} & =\frac{\mathcal{A}^{11}}{\sqrt{\mathcal{A}}}\nonumber \\
\kappa_{12} & =-\frac{\mathcal{A}^{12}}{\sqrt{\mathcal{A}}}.
\label{338}
\end{align}
\end{equation}
It is easy to see with a direct calculation that this solution satisfies 
the constraint $\kappa=1$.\\
We can now elaborate a bit the solution (\ref{338}). We can write

\[ \left[
\begin{array}{ll}
\kappa_{11} & \kappa_{12}  \\    
\kappa_{21} & \kappa_{22}
\end{array} 
\right] =
\left[ 
\begin{array}{cc}
\mathcal{A}^{22}/\sqrt{\mathcal{A}} & -\mathcal{A}^{12}/\sqrt{\mathcal{A}} \\
-\mathcal{A}^{21}/\sqrt{\mathcal{A}} & \mathcal{A}^{11}/\sqrt{\mathcal{A}}
\end{array}
\right] =   
\sqrt{\mathcal{A}}\sqrt{\mathcal{A}}
\left[
\begin{array}{cc}
\mathcal{A}^{22}/\mathcal{A} & -\mathcal{A}^{12}/\mathcal{A} \\
-\mathcal{A}^{21}/\mathcal{A} & \mathcal{A}^{11}/\mathcal{A}
\end{array}
\right] \] \\
\[ = \mathcal{A}[\mathcal{A}^{AB}]^{-1} =
\mathcal{A}[\mathcal{A}_{AB}] =
\left[
\begin{array}{ll}
\sqrt{\mathcal{A}}\mathcal{A}_{11} & \sqrt{\mathcal{A}}\mathcal{A}_{12} \\
\sqrt{\mathcal{A}}\mathcal{A}_{21} & \sqrt{\mathcal{A}}\mathcal{A}_{22}
\end{array}
\right].   
\] 
This means 
\begin{equation}
\kappa_{AB}=\sqrt{\mathcal{A}}\mathcal{A}_{AB}
\end{equation}
where $\mathcal{A}=\det\mathcal{A}^{AB}$. Therefore 
$\det\mathcal{A}_{AB}=1/\mathcal{A}$ and
\begin{equation}
\kappa^{AB}=\frac{1}{\sqrt{\mathcal{A}}}\mathcal{A}^{AB}.
\end{equation}
In particular we have again
\begin{equation}
\det\kappa_{AB} = \det(\sqrt{\mathcal{A}}\mathcal{A}_{AB})
=\mathcal{A}\det\mathcal{A}_{AB}=1.
\end{equation}
The expressions for the solutions of the $\lambda$-C are
\begin{equation}
\lambda_{AB}=\sqrt{\lambda}\;\kappa_{AB}=\sqrt{\lambda\mathcal{A}}\;\mathcal{A}_{AB}\quad;
\quad \quad \quad \lambda^{AB}= \frac{1}{\sqrt{\lambda\mathcal{A}}}\mathcal{A}^{AB}.
\label{sollambdaC}
\end{equation}

\section{Substitution of $\lambda_{AB}$ (solution of the $\lambda$-C) 
in the shell part of $\mathcal{H}_{118}^{NEW(2)}$}

In this section we substitute back into the shell part of $\mathcal{H}_{118}^{NEW(2)}$
the solution $\lambda_{AB}\equiv q_{AB}|_{\Sigma}$ of the $\lambda$-C, constructed 
in the previous section.
The substitution has to be done in the shell part only, because $\lambda_{AB}$ 
is defined only on the shell $S\cap\Sigma$. \\
The shell part of $\mathcal{H}_{118}^{NEW(2)}$, relevant for the substitution, is
\begin{equation}
\begin{align}
\mathcal{H}_{118}^{NEW(2)}(SHELL)  =
& -\frac{1}{8 \pi G} \int_{S\cap\Sigma}d^{2}\eta
\sqrt{\lambda}([\tilde{\pi}^{\perp\perp}N^{\perp}-lN] \nonumber \\
& +\lambda^{KL}N_{L}
([\tilde{\pi}_{K}^{\perp}]+E\left(\frac{N_{-}^{\perp}}{N_{-}}\right)_{,K})) 
- \int_{S\cap\Sigma}d^{2}\eta T^{0}_{s 0} 
\end{align}
\end{equation}
Everywhere it is understood that $N_{+}^{\perp}$, $N_{+}$ are given by (\ref{N+N-}).
Note that $\lambda_{AB}$ appears in $l=l_{AB}\lambda^{AB}$, in 
$T^{0}_{s 0}(\nu, \nu^{K}, \lambda_{KL}, z^{A}, p_{A})$ and in $\lambda^{KL}N_{L}$.
With the substitutions
\begin{equation}
\begin{align}
\lambda_{AB} & =\sqrt{\lambda\mathcal{A}}\;\mathcal{A}_{AB}\nonumber \\
\lambda^{AB} & = \frac{1}{\sqrt{\lambda\mathcal{A}}}\mathcal{A}^{AB}
\end{align}
\end{equation}
where
\begin{equation}
\mathcal{A}^{AB}:=\sqrt{\lambda}[\tilde{\pi}^{AB}N^{\perp}+Nl^{AB}]+8 \pi G T^{AB}
\end{equation}
the shell part of the hamiltonian becomes
\begin{equation}
\begin{align}
\mathcal{H}_{118}^{NEW(2)} & (SHELL)=
-\frac{1}{8 \pi G} \int_{S\cap\Sigma}d^{2}\eta
\sqrt{\lambda}([\tilde{\pi}^{\perp\perp}N^{\perp}-
Nl_{AB}(\frac{1}{\sqrt{\lambda\mathcal{A}}}\mathcal{A}^{AB})]\nonumber \\
& +\frac{1}{\sqrt{\lambda\mathcal{A}}}\mathcal{A}^{KL}N_{L}
([\tilde{\pi}_{K}^{\perp}]+E\left(\frac{N_{-}^{\perp}}{N_{-}}\right)_{,K})) 
- \int_{S\cap\Sigma}d^{2}\eta T^{0}_{s 0}
\end{align}
\end{equation}
We remind again here that $N_{+}^{\perp}$, $N_{+}$ are as in (\ref{N+N-}).
We see now that the variables $\phi$, $\chi$ have disappeared from this hamiltonian
and therefore the variation of it in respect to them is equal to zero. 
So the singular constraint $\lambda$-C is automatically satisfied and it does not 
appear anymore in the final equations. 

\section{Boundary conditions for $q_{AB}$ at $\Sigma$}

We have seen  that in adapted coordinates the $2+1$ version of the 
$\gamma$-C holds
\begin{equation}
(q_{kl}e^{k}_{A}e^{l}_{B})^{\pm}=q_{AB}|_{S\cap\Sigma}=\lambda_{AB}.
\end{equation}
The $\pm$ signs mean that we are taking the limits of $q_{kl}e^{k}_{A}e^{l}_{B}$
from $S^{\pm}$ towards $S\cap\Sigma$. Now the expression of $\lambda_{AB}$
cannot be fixed arbitrarily "by hand" as before, but it is given as the solution of the 
$\lambda$-C. Therefore the $\gamma$-C can now be interpreted as a boundary 
condition for $q_{AB}$ at $\Sigma$. Precisely
\begin{equation}
(q_{kl}e^{k}_{A}e^{l}_{B})\stackrel{\Sigma}{\longrightarrow}
\lambda_{AB}=\sqrt{\lambda\mathcal{A}}\mathcal{A}_{AB}.
\end{equation}
In other words, the metric $q_{kl}$ has to satisfy this condition at the 
boundary $S\cap\Sigma$, namely
\begin{equation}
(q_{kl}e^{k}_{A}e^{l}_{B})^{\pm}=q_{AB}|_{\Sigma}=\lambda_{AB}=
\sqrt{\lambda\mathcal{A}}\;\mathcal{A}_{AB}.
\end{equation}
where
\begin{equation}
\mathcal{A}^{AB}:=\sqrt{\lambda}[\tilde{\pi}^{AB}N^{\perp}+Nl^{AB}]+8 \pi G T^{AB}.
\end{equation} 
and of course $N_{+}^{\perp}=N_{+}^{\perp}(N_{-}, N_{-}^{\perp})$,
$N_{+}=N_{+}(N_{-}, N_{-}^{\perp})$. 
These conditions are analogous to fall-off conditions at infinity. 
They define the phase space of our system, in the same way as fall-off 
conditions do.\\
After the solution of the $\lambda$-C, it is clear that the $S\cap\Sigma$-variables
now independent are $N_{-}$, $N_{-}^{\perp}$, $N_{1}$, $N_{2}$, $\lambda$.
The variables $\alpha_{\pm}$, $[\alpha]$, $\nu$, $\nu^{K}$, $\lambda_{KL}$, 
$q_{KL}|_{\Sigma}$, $N_{+}^{\perp}$, $N_{+}$ can be thought as \textit{auxiliary
variables}, and they can be expressed as functions of the former. We have already 
seen these relations in the previous sections; the last one here obtained is 
\begin{equation}
\lambda_{AB}=q_{AB}|_{\Sigma}=\sqrt{\lambda\mathcal{A}}\;\mathcal{A}_{AB}.
\end{equation}   
\underline{The new hamiltonian}, after the solution of the $\lambda$-C, is
\begin{equation}
\begin{align}
\mathcal{H}_{118}^{NEW(3)} & =
\frac{1}{16 \pi G} \int_{S^{-}\cup S^{+}}d^{3}y\{NC+N^{k}C_{k}\} \nonumber \\ 
& -\frac{1}{8 \pi G} \int_{S\cap\Sigma}d^{2}\eta
\sqrt{\lambda}\{[\tilde{\pi}^{\perp\perp}N^{\perp}-
Nl_{AB}(\frac{1}{\sqrt{\lambda\mathcal{A}}}\mathcal{A}^{AB})]\nonumber \\
& +\frac{1}{\sqrt{\lambda\mathcal{A}}}\mathcal{A}^{KL}N_{L}
([\tilde{\pi}_{K}^{\perp}]+E\left(\frac{N_{-}^{\perp}}{N_{-}}\right)_{,K})\} 
- \int_{S\cap\Sigma}d^{2}\eta T^{0}_{s 0}
\end{align}
\end{equation}
\underline{The final phase space} is described by the canonical variables\\ \\
\( \begin{array}{cll}
\hspace*{3cm}-\frac{1}{16 \pi G}\;q_{kl} & \longleftrightarrow & \pi^{kl} \\ \\
\hspace*{3cm}-\frac{1}{8 \pi G}\sqrt{\lambda} & \longleftrightarrow & 
E\left(\frac{N_{-}^{\perp}}{N_{-}}\right) \\ \\
\hspace*{3cm}z^{A} & \longleftrightarrow & p_{A}
\end{array} \) \\ \\  
\underline{The final symplectic form} can be written as
\begin{equation}
\begin{align}
\delta\mathcal{H}^{SYM(3)}_{107(NEW)} & = \frac{1}{16 \pi G}\int_{S}d^{3}y(-\dot{\pi}^{kl}
\delta q_{kl}+\dot{q}_{kl}\delta\pi^{kl}) \nonumber \\
& +\frac{1}{16 \pi G}\int_{S\cap\Sigma}d^{2}
\eta\sqrt{\lambda}  
\left(-\dot{E}\left(\frac{N_{-}^{\perp}}{N_{-}}\right)\lambda^{KL}\delta\lambda_{KL}
+\frac{\dot{\lambda}}{\lambda}\delta E\left(\frac{N_{-}^{\perp}}{N_{-}}\right)\right) \nonumber \\ 
& -\int_{S\cap\Sigma}d^{2}\eta(\dot{p}_{A}\delta
z^{A}-\dot{z}^{A}\delta p_{A}).
\end{align}
\end{equation}
Also from here we can read the phase space variables.\\
\underline{The final variation of $\mathcal{H}_{118}^{NEW(3)}$} 
can be calculated in the usual way, using the auxiliary variables when opportune,
and the rules of variational calculus. Reminding that now $N_{+}, N^{\perp}_{+}$
are expressed by $N_{-}, N^{\perp}_{-}$, we get an expression formally identical
to (\ref{varH118N2})
\begin{equation}
\begin{align}
\delta\mathcal{H}_{118}^{NEW(3)} & =\frac{1}{16 \pi G}\int_{S}d^{3}y(C_{k}\delta N^{k}+
C\delta N + a^{kl}\delta q_{kl} + b_{kl}\delta\pi^{kl}) \nonumber \\
& +\frac{1}{16 \pi G}
\int_{S\cap\Sigma}d^{2}\eta[2REST-\sqrt{\lambda}B^{k}m_{k}] 
- \int_{S\cap\Sigma}d^{2}\eta\left\{\frac{1}{2}T^{KL}_{s}\delta\lambda_{KL}\right\} \nonumber \\
& -\frac{1}{8 \pi G}\int_{S\cap\Sigma}d^{2}\eta\frac{\sqrt{\lambda}}
{\sqrt{|\gamma|}}
\{\left([Q^{\perp\perp}]-8\pi G T^{\perp\perp}_{s}\right)
\delta\left(\sqrt{(N_{-})^{2}-(N_{-}^{\perp})^{2}}\right) \nonumber \\
& +\left([Q^{\perp}_{K}]-8 \pi GT^{\perp}_{s\;K}\right)\delta(N^{K})\}
\nonumber \\
& - \int_{S\cap\Sigma}d^{2}\eta\left\{\left(\frac{\partial T^{0}_{s\;0}}
{\partial z^{A}}-\frac{\partial}{\partial\eta^{M}}\frac{\partial T^{0}_{s\;0}}
{\partial z^{A}_{M}}\right)\delta z^{A}+ \frac{\partial T^{0}_{s\;0}}{\partial 
p_{A}}\delta p_{A}\right\}.
\end{align}
\end{equation} \\
but where $\lambda_{KL}=\sqrt{\lambda\mathcal{A}}\;\mathcal{A}_{KL}$.\\
\underline{Final equations and constraints}: The comparison
\begin{equation}
\delta\mathcal{H}_{118}^{NEW(3)}=\delta\mathcal{H}^{SYM(3)}_{107(NEW)}. 
\end{equation}
gives the standard canonical equations already stated after eq.(\ref{SCE}), 
again unchanged in form
\[ \left\{ \begin{array}{ll}
C_{k}=0 \\ \\
C=0
\end{array}
\right. \quad \left\{ \begin{array}{ll}
\dot{q}_{kl}=b_{kl} \\ \\
\dot{\pi}^{kl}=-a^{kl}
\end{array}
\right.  \quad \left\{ \begin{array}{ll} 
[Q^{\perp\perp}]=8 \pi G T^{\perp\perp}_{s} \\ \\

[Q^{\perp}_{K}]= 8 \pi G T^{\perp}_{s\;K}
\end{array}
\right.    \quad \left\{ \begin{array}{ll}
\dot{p}_{A}=\frac{\partial T^{0}_{s\;0}}
{\partial z^{A}}-\frac{\partial}{\partial\eta^{M}}\frac{\partial T^{0}_{s\;0}}
{\partial z^{A}_{M}} \\ \\
-\dot{z}^{A}=\frac{\partial T^{0}_{s\;0}}{\partial p_{A}}
\end{array}
\right.    \] \\ \\
The remaining "REST" terms, using the variation (\ref{varH118rest}), give
\begin{equation}
\begin{align}
& \int_{S\cap\Sigma}d^{2}\eta\{-\sqrt{\lambda}([\tilde{\pi}^{KL}
N^{\perp}+l^{KL}N-l\lambda^{KL}N-N_{,k}m^{k}\lambda^{KL}] \nonumber \\
& +\lambda^{MR}N_{R}\lambda^{KL}E(\frac{N_{-}^{\perp}}{N_{-}})_{,M})\delta\lambda_{KL}
+2\sqrt{\lambda}(-\tilde{\pi}^{\perp\perp}N+lN^{\perp}
+(\lambda^{KL}N_{L})_{||K})^{\pm}\delta E(\frac{N_{-}^{\perp}}{N_{-}})\} \nonumber \\
& -\int_{S\cap\Sigma}d^{2}\eta(8 \pi G T^{KL}_{s}\delta\lambda_{KL}) \nonumber \\
& =\int_{S\cap\Sigma}d^{2}
\eta\sqrt{\lambda}\left\{-\dot{E}\left(\frac{N_{-}^{\perp}}{N_{-}}\right)\lambda^{KL}\delta\lambda_{KL}
+\frac{\dot{\lambda}}{\lambda}\delta E\left(\frac{N_{-}^{\perp}}{N_{-}}\right)\right\}
\end{align}
\end{equation}  
where the variables $N_{+}, N^{\perp}_{+}, N_{-}, N^{\perp}_{-}$ satisfy by 
construction the $\alpha$-C and $\lambda_{KL}=\sqrt{\lambda\mathcal{A}}\;\mathcal{A}_{KL}$
is now the solution of the $\lambda$-C. 
From the comparison we get the equations:\\ \\
1)- The \underline{final form} of the equation of motion for $\lambda$, 
(eq.131HK or eq.(\ref{lambda}))
\begin{equation}
\dot{\lambda}=2\lambda(-\tilde{\pi}^{\perp\perp}N+l_{AB}\lambda^{AB}N^{\perp}
+(\lambda^{KL}N_{L})_{||K})^{\pm}
\end{equation}
where of course $N_{+}$, $N_{+}^{\perp}$ are given by (\ref{N+N-}) and
$\lambda^{KL}=(\lambda\mathcal{A})^{-1/2}\mathcal{A}^{KL}$.\\ \\
2)- The \underline{final form} of the equation 130HK (eq.\ref{new130HK2})
\begin{equation}
\begin{align}
& -\sqrt{\lambda}([\tilde{\pi}^{KL}
N^{\perp}+l^{KL}N-l\lambda^{KL}N-N_{,k}m^{k}\lambda^{KL}]
+\lambda^{MR}N_{R}\lambda^{KL}E(\frac{N_{-}^{\perp}}{N_{-}})_{,M}) \nonumber \\
& -8 \pi G T^{KL}_{s}= 
-\sqrt{\lambda}\lambda^{KL}\dot{E}\left(\frac{N_{-}^{\perp}}{N_{-}}\right) 
\label{fin130HK}
\end{align}
\end{equation}
where now $\lambda^{KL}$ is an abreviation for 
$(\lambda\mathcal{A})^{-1/2}\mathcal{A}^{KL}$ and $N_{+}$, $N_{+}^{\perp}$ 
are as before.\\ \\
The trace of (\ref{fin130HK}) is the final form of the equation of motion 137HK,
i.e. an equation of motion for the variable $E(N_{-}^{\perp}/N_{-})$
\begin{equation}
\begin{align}
\sqrt{\lambda}\dot{E}\left(\frac{N_{-}^{\perp}}{N_{-}}\right)
= 4 \pi G T^{KL}\lambda_{KL}
& +\frac{1}{2}\sqrt{\lambda}([\tilde{\pi}^{KL}\lambda_{KL}N^{\perp}-lN-2N_{,k}m^{k}] \nonumber \\
& +2\lambda^{MR}N_{R}E(\frac{N_{-}^{\perp}}{N_{-}})_{,M}) 
\end{align}
\end{equation} 
where the $\lambda$'s are given by eqs.(\ref{sollambdaC}) and $N_{+}$, $N_{+}^{\perp}$ 
are as before in (\ref{N+N-}).\\ \\
The trace-free part of (\ref{fin130HK}) is the $\lambda$-C: but now, in the new 
variables, it is (by construction) identically satisfied and equal to zero.\\ \\
We see that the singular constraints $\alpha$-C and $\lambda$-C have been eliminated
from the final equations. In the end we have been left only with the equations of 
motion for the variables $\lambda$ and $E(N_{-}^{\perp}/N_{-})$, plus, of course,
canonical constraints.
It is important to note that these conclusions are in full agreement with the general
scheme of the theorem stated in $\S$ \ref{GT}. 
%
%
%
%
%
%
%
%
%
\chapter{Differentiability}
\section{Definition of differentiability (in our context)}
In this section we define what we mean for differentiability of a functional 
in the context of our work. We do not adopt the very general definition coming from
the manuals of Functional Analysis, because such definition would require the 
introduction of a norm and a topology in our phase space, and this procedure is too 
long and complicated for our scopes.\\
Indeed we adopt a very simple operative definition of differentiability, adapted to 
our geometrical context:\\ \\
\textbf{Definition} $\;$ \textit{A given functional $\mathcal{H}(q(y), u(\eta))$, 
defined on our geometrical 
environment $S$, $S\cap\Sigma$, is said to be "differentiable" if its total 
variation can be written as 
\begin{equation}
\delta\mathcal{H}= \int_{S}d^{3}y F(q(y))\delta q(y) + \int_{S\cap\Sigma}d^{2}\eta
G(u(\eta))\delta u(\eta)
\end{equation}
where $\delta q(y)$ must be a function of 3 variables $y^{1},y^{2},y^{3}$
defined on $S$, and $\delta u(\eta)$ must be a function of 2 variables
$\eta^{1}, \eta^{2}$ defined on $S \cap \Sigma$}.\\ \\
This definition is useful for our scope: if we want to calculate, for example, 
a Poisson bracket, we have to consider terms as 
\begin{equation}
\frac{\delta\mathcal{H}}{\delta q(y')}; \quad \quad \quad 
\frac{\delta\mathcal{H}}{\delta u(\eta')}.
\end{equation}
Such terms produce Dirac $\delta$ functions
\begin{equation}
\frac{\delta q(y)}{\delta q(y')}=\delta^{3}(y, y'); \quad \quad \quad 
\frac{\delta u(\eta)}{\delta u(\eta')}=\delta^{2}(\eta, \eta').
\end{equation}
And these $\delta$'s must be of the correct dimension in order to generate 
numbers after the integration. For example, we have
\begin{multline}
\frac{\delta\mathcal{H}}{\delta q(y')}=\int_{S}d^{3}y F(q(y))\frac{\delta q(y)}{\delta q(y')} \\
=\int_{S}d^{3}y F(q(y))\delta^{3}(y, y')=F(q(y')) \equiv \; NUMBER
\end{multline}
and the analog for $\eta$. It is clear that an object like
\begin{equation}
\int_{S\cap\Sigma}d^{2}y F(q(y))\delta q(y)
\end{equation}
is not admitted, because it would generate a not-absorbable Dirac $\delta$:
\begin{multline}
\int_{S\cap\Sigma}d^{2}y F(q(y))\frac{\delta q(y)}{\delta q(y')}=
\int_{S\cap\Sigma}d^{2}y F(q(y))\delta^{3}(y, y') \\
=F(q(y_{1}',y_{2}'))\delta(y_{3}, y_{3}') \equiv NOT \; A \; NUMBER \; !
\end{multline}
Therefore the differentiability of a functional $\mathcal{F}$ means that 
$\delta\mathcal{F}/\delta q$ is a number and not a $\delta$-function.

\section{Differentiability of the initial hamiltonian}

We will see in this section as the initial hamiltonian $\mathcal{H}_{118}$
of our system is a differentiable functional in adapted coordinates.
In fact, if we describe our system using adapted coordinates, we know that 
the following equations hold
\begin{equation}
\begin{align}
[N^{\perp}-\nu\sinh\alpha]=0 \nonumber \\
[N-\nu\cosh\alpha]=0.
\end{align}
\end{equation}
These two relations (continuity relations), between surface and volume 
smearing functions, guarantee an effective differentiability of the whole
hamiltonian $\mathcal{H}_{118}$. To see this, it is sufficient to write 
down the total variation of $\mathcal{H}_{118}$. Reminding (\ref{H118}) and
(\ref{1.12}), we can write
\begin{equation}
\begin{align}
\delta\mathcal{H}_{118} & =\frac{1}{16 \pi G}\int_{S}d^{3}y(C_{k}\delta N^{k}+
C\delta N + a^{kl}\delta q_{kl} + b_{kl}\delta\pi^{kl}) \nonumber \\
& +\frac{1}{16 \pi G}
\int_{S\cap\Sigma}d^{2}\eta[2REST-\sqrt{\lambda}B^{k}m_{k}] 
- \int_{S\cap\Sigma}d^{2}\eta\left\{\frac{1}{2}T^{KL}_{s}\delta\lambda_{KL}\right\} 
\nonumber \\
& -\frac{1}{8 \pi G}\int_{S\cap\Sigma}d^{2}\eta\frac{\sqrt{\lambda}}
{\sqrt{|\gamma|}}\left\{\left([Q^{\perp\perp}]-8\pi G T^{\perp\perp}_{s}\right)\delta\nu+
\left([Q^{\perp}_{K}]-8 \pi GT^{\perp}_{s\;K}\right)\delta\nu^{K}\right\} \nonumber \\ 
& - \int_{S\cap\Sigma}d^{2}\eta\left\{\left(\frac{\partial T^{0}_{s\;0}}
{\partial z^{A}}-\frac{\partial}{\partial\eta^{M}}\frac{\partial T^{0}_{s\;0}}
{\partial z^{A}_{M}} \right)\delta z^{A}+ \frac{\partial T^{0}_{s\;0}}{\partial 
p_{A}}\delta p_{A}\right\}
\label{*1}
\end{align}
\end{equation}
and
\begin{equation}
\begin{align}
& \int_{S\cap\Sigma}d^{2}\eta[2REST-\sqrt{\lambda}B^{k}m_{k}]= \nonumber \\ 
= & \int_{S\cap\Sigma}d^{2}\eta[-2\sqrt{\lambda}\tilde{\pi}^{\perp\perp}
(\nu\sinh\alpha-N^{\perp})m^{k}m^{l}\delta q_{kl} - \sqrt{\lambda}
\{\tilde{\pi}^{\perp\perp}\lambda^{KL}(\nu\sinh\alpha-N^{\perp}) \nonumber \\
& +\tilde{\pi}^{KL}
N^{\perp}+l^{KL}N-l\lambda^{KL}\nu\cosh\alpha+\nu^{M}\lambda^{KL}\alpha_{,M}
-N_{,k}m^{k}\lambda^{KL}\}\delta\lambda_{KL} \nonumber \\
& +2\sqrt{\lambda}(\nu\cosh\alpha-N)\delta l- 
2\sqrt{\lambda}(\nu\sinh\alpha-N^{\perp})m_{k}m_{l}\delta\tilde{\pi}^{kl}] \nonumber \\
& +2\sqrt{\lambda}[-\nu\tilde{\pi}^{\perp\perp}\cosh\alpha+l\nu\sinh\alpha
+\nu^{K}_{||K}]\delta\alpha^{\pm} \nonumber \\
& +2\sqrt{\lambda}(-\nu\tilde{\pi}^{\perp\perp}\cosh\alpha+l\nu\sinh\alpha
+\nu^{K}_{||K})^{\pm}\delta[\alpha]. 
\label{**1}
\end{align}
\end{equation}
Inspecting the numerous terms, we see that in (\ref{*1}) there is always the correct 
agreement between the dimension of the measure and the dimension of the variation.
In particular $\delta N^{k}$, $\delta N$, $\delta q_{kl}$, $\delta \pi^{kl}$ depend
on 3 variables $y^{1}$, $y^{2}$, $y^{3}$, while $\delta\lambda_{KL}$, $\delta\nu$,
$\delta\nu^{K}$, $\delta z^{A}$, $\delta p_{A}$ depend on 2 variables $\eta^{1}$,    
$\eta^{1}$. \\
Yet, there are still problems with the following terms contained in "REST" (\ref{**1}):
\begin{equation}
\begin{align}
& -2\sqrt{\lambda}\tilde{\pi}^{\perp\perp}
(\nu\sinh\alpha-N^{\perp})m^{k}m^{l}\delta q_{kl} \nonumber \\
& +2\sqrt{\lambda}(\nu\cosh\alpha-N)\delta l \nonumber \\
& -2\sqrt{\lambda}(\nu\sinh\alpha-N^{\perp})m_{k}m_{l}\delta\tilde{\pi}^{kl}. \nonumber 
\end{align}
\end{equation}
In fact, $\delta q_{kl}$, $\delta\tilde{\pi}^{kl}$, $\delta l \equiv \delta(q^{kl}m_{k|l})$
depend on 3 variables but they are integrated only over 
$\int_{S\cap\Sigma}d^{2}\eta$. Therefore they could generate $\delta$'s. 
But, luckly enough, we are in adapted coordinates, and therefore all those "bad" terms
cancel out. In other words, we see here how the role of the "continuity constraint",
i.e. the $\gamma$-C, (which requires the introduction of adapted coordinates) is 
fundamental in order to guarantee the differentiability of the hamiltonian.

\section{Differentiability of the final hamiltonian}

After the solution of the $\gamma$-C, $\alpha$-C, $\lambda$-C and the complete 
reduction of the system, the final hamiltonian reads
\begin{equation}
\begin{align}
\mathcal{H}_{118}^{NEW(3)} & =
\frac{1}{16 \pi G} \int_{S^{-}\cup S^{+}}d^{3}y\{NC+N^{k}C_{k}\} \nonumber \\ 
& -\frac{1}{8 \pi G} \int_{S\cap\Sigma}d^{2}\eta
\sqrt{\lambda}\{[\tilde{\pi}^{\perp\perp}N^{\perp}-Nl]\nonumber \\
& +N^{K}([\tilde{\pi}_{K}^{\perp}]+E\left(\frac{N_{-}^{\perp}}{N_{-}}\right)_{,K})\} 
- \int_{S\cap\Sigma}d^{2}\eta T^{0}_{s 0}
\end{align}
\end{equation}
where in particular $\lambda_{KL}=\sqrt{\lambda\mathcal{A}}\;\mathcal{A}_{KL}$,
$l=(\lambda\mathcal{A})^{-1/2}l_{AB}\mathcal{A}^{AB}$, \\
$N^{K}=(\lambda\mathcal{A})^{-1/2}\mathcal{A}^{KL}N_{L}$.
To check its differentiability, let's write down its total variation
\begin{equation}
\begin{align}
\delta\mathcal{H}_{118}^{NEW(3)} & =\frac{1}{16 \pi G}\int_{S}d^{3}y(C_{k}\delta N^{k}+
C\delta N + a^{kl}\delta q_{kl} + b_{kl}\delta\pi^{kl}) \nonumber \\
& +\frac{1}{16 \pi G}
\int_{S\cap\Sigma}d^{2}\eta[2REST-\sqrt{\lambda}B^{k}m_{k}] 
- \int_{S\cap\Sigma}d^{2}\eta\left\{\frac{1}{2}T^{KL}_{s}\delta\lambda_{KL}\right\} \nonumber \\
& -\frac{1}{8 \pi G}\int_{S\cap\Sigma}d^{2}\eta\frac{\sqrt{\lambda}}
{\sqrt{|\gamma|}}
\{\left([Q^{\perp\perp}]-8\pi G T^{\perp\perp}_{s}\right)
\delta\left(\sqrt{(N_{-})^{2}-(N_{-}^{\perp})^{2}}\right) \nonumber \\
& +\left([Q^{\perp}_{K}]-8 \pi GT^{\perp}_{s\;K}\right)\delta(N^{K})\}
\nonumber \\
& - \int_{S\cap\Sigma}d^{2}\eta\left\{\left(\frac{\partial T^{0}_{s\;0}}
{\partial z^{A}}-\frac{\partial}{\partial\eta^{M}}\frac{\partial T^{0}_{s\;0}}
{\partial z^{A}_{M}}\right)\delta z^{A}+ \frac{\partial T^{0}_{s\;0}}{\partial 
p_{A}}\delta p_{A}\right\}.
\end{align}
\end{equation}
Now, inspect accurately this variation.\\
The first line does not create problems
because $\delta N^{k}$, $\delta N$, $\delta q_{kl}$, $\delta \pi^{kl}$ are functions 
of 3 variables ($y^{1}$, $y^{2}$, $y^{3}$) integrated over $\int_{S}d^{3}y$.\\ 
The second line will be examinated after.\\
The third and fourth line contain $N_{-}$, $N_{-}^{\perp}$, $N^{K}$. 
It is clear from preceding considerations (see beginning of chapter 3) that all 
these variables are explicitly defined only on the shell $S\cap\Sigma$. They depend 
only on $\eta^{1}$, $\eta^{2}$. Hence we have functions of two variables 
integrated over $d^{2}\eta$, so all the $\delta$'s can be re-absorbed and there are 
no problems.\\
The same can be said about the last line. The variables $z^{A}$, $p_{A}$ are defined on 
$S\cap\Sigma$ and they depend only on $\eta^{1}$, $\eta^{2}$ 
(apart from $t$, which is a parameter). So also here the $\delta$'s can be re-absorbed
by the integral in $d^{2}\eta$.\\
Let's now examine the second line. Explicitly, it reads
\begin{multline}
\int_{S\cap\Sigma}d^{2}\eta\{-\sqrt{\lambda}([\tilde{\pi}^{KL}
N^{\perp}+l^{KL}N-l\lambda^{KL}N-N_{,k}m^{k}\lambda^{KL}] \\
+N^{M}\lambda^{KL}E(\frac{N_{-}^{\perp}}{N_{-}})_{,M})\delta\lambda_{KL}
 +2\sqrt{\lambda}(-\tilde{\pi}^{\perp\perp}N+lN^{\perp}
+(\lambda^{KL}N_{L})_{||K})^{\pm}\delta E(\frac{N_{-}^{\perp}}{N_{-}})\} \\
-\int_{S\cap\Sigma}d^{2}\eta(8 \pi G T^{KL}_{s})\delta\lambda_{KL}.
\end{multline}
We remind now that 
\begin{equation}
E\left(\frac{N_{-}^{\perp}}{N_{-}}\right) \equiv [\alpha].
\end{equation}
The variables $[\alpha]$, $\alpha^{\pm}$ (as well as $N_{-}^{\perp}$, $N_{-}$) are 
defined only on the shell $S\cap\Sigma$. This comes directly from the definition of
$N_{-}^{\perp}$, $N_{-}$, $\alpha^{\pm}$. Therefore $E(N_{-}^{\perp}/N_{-})$ depends
only on $\eta^{1}$, $\eta^{2}$. So we have a function of 2 variables integrated over 
$d^{2}\eta$ and therefore all possible $\delta$'s can be re-absorbed by the integral.\\
We also remind that $\lambda_{KL}=\sqrt{\lambda\mathcal{A}}\;\mathcal{A}_{KL}$ with
\begin{equation}
\mathcal{A}^{KL}:=\sqrt{\lambda}[\tilde{\pi}^{KL}N^{\perp}+Nl^{KL}]+8 \pi G T^{KL}.
\end{equation} 
The variable $\lambda_{KL}$ is explicitly defined \textit{only} on $S\cap\Sigma$.
In particular, from the $\gamma$-C we have 
\begin{equation}
\lambda_{KL}=q_{KL}|_{S\cap\Sigma}.
\end{equation}
Moreover in the definition of $\mathcal{A}^{KL}$ appears a jump [...] which forces
the variables $N_{-}^{\perp}$, $N_{-}$ to be calculated at $S\cap\Sigma$. And the same
can be said about the variables $\tilde{\pi}^{KL}$, $l^{KL}$, which are volume quantities,
but explicitly projected and calculated \underline{on} $S\cap\Sigma$. Also 
$T^{KL}_{s}$ is the energy-momentum tensor of the shell calculated at $S\cap\Sigma$. 
In other words, the field $\lambda_{KL}$, which is the metric on the shell $S\cap\Sigma$,
cannot be thought depending on variables other than $\eta^{1}$, $\eta^{2}$. 
Therefore also here the $\delta$'s can be re-absorbed by the integral 
$\int_{S\cap\Sigma}d^{2}\eta$. \\
So, we conclude that the final hamiltonian $\mathcal{H}_{118}^{NEW(3)}$ is a 
differentiable functional. The role of the $\gamma$-C and the adapted coordinates 
in guaranteeing the differentiabilty is here clearly evident.

\section{Differentiability of the final constraints}

We note that the initial hamiltonian can be written as sum of canonical volume and
surface constraints. In fact, reminding (\ref{H108}), we can write
\begin{equation}
\mathcal{H} = \frac{1}{8 \pi G} \int_{S^{-}\cup S^{+}}d^{3}y G^{0}_{0} +
\frac{1}{8 \pi G} \int_{S\cap\Sigma}d^{2}\eta[Q_{0}^{0}]-\int_{S\cap\Sigma}
d^{2}\eta T^{0}_{s\;0}.
\end{equation}
Using now the well known decompositions (see HK paper formulae 112.4, 112.7, 112.8)
\begin{equation}
\begin{align}
G^{0}_{0} & =\frac{1}{2}(NC+N^{k}C_{k}) \nonumber \\
[Q_{0}^{0}] & = -\frac{\sqrt{\lambda}}{\sqrt{|\gamma|}}(\nu[Q^{\perp\perp}]+
\nu^{K}[Q^{\perp}_{K}]) \nonumber \\
T^{0}_{s\;0} & = -\frac{\sqrt{\lambda}}{\sqrt{|\gamma|}}(\nu T^{\perp\perp}_{s}+
\nu^{K}T^{\perp}_{s\;K})
\end{align}
\end{equation}
we can write
\begin{equation}
\begin{align}
\mathcal{H}_{118} & =\frac{1}{16 \pi G}\int_{S}d^{3}y(N^{k}C_{k} +NC) \nonumber \\
& -\frac{1}{8 \pi G}\int_{S\cap\Sigma}d^{2}\eta\frac{\sqrt{\lambda}}
{\sqrt{|\gamma|}}\left\{\nu\left([Q^{\perp\perp}]-8\pi G T^{\perp\perp}_{s}\right)+
\nu^{K}\left([Q^{\perp}_{K}]-8 \pi GT^{\perp}_{s\;K}\right)\right\}.
\end{align}
\end{equation}  
We have shown that $\mathcal{H}_{118}$ is a differentiable functional on the phase space.
Bacause of the arbitrariness of the lagrangian multipliers, the differentiability
of $\mathcal{H}_{118}$ implies directly the differentiability of the constraints 
(in fact, we can set alternatively the multipliers equal to zero, or to fixed values). 
Therefore we can say that the canonical initial constraints are differentiable 
functionals. \\ \\
The same argument can be used with the final canonical constraints. 
In fact, the final hamiltonian can be written as the sum of 
the canonical (volume and surface) constraints 
\begin{multline}
\mathcal{H}_{118}^{NEW(3)}=
\frac{1}{16 \pi G}\int_{S}d^{3}y(N^{k}C_{k} +NC) \\
-\frac{1}{8 \pi G}\int_{S\cap\Sigma}d^{2}\eta\frac{\sqrt{\lambda}}
{\sqrt{|\gamma|}}\{\sqrt{(N_{-})^{2}-(N_{-}^{\perp})^{2}}
\left([Q^{\perp\perp}]-8\pi G T^{\perp\perp}_{s}\right) \\
+N^{K}\left([Q^{\perp}_{K}]-8 \pi GT^{\perp}_{s\;K}\right)\}.
\end{multline}
So, because the final hamiltonian \textit{is} 
differentiable and it is a linear combination of the canonical constraints
(via arbitrary lagrangian multipliers),
we can say that also the final canonical constraints (volume and surface) 
are differentiable functionals on the final phase space.
Therefore their Poisson brackets can be computed and are well defined objects.
%
%
%
%
%
%
%
%
\chapter{Conclusions}
In this work we have showed that the singular constraints $\gamma$-C, 
$\alpha$-C, $\lambda$-C, which plagued the hamiltonian system described in 
the HK paper ~\cite{HK}, can be eliminated with a progressive reduction 
procedure ~\cite{Teitel}.\\
The variables describing the final phase space are clearly identified.
The final hamiltonian and the final symplectic form are expressed by means 
of these variables alone. The final form of equations of motion and 
constraints is explicitly stated. The only constraints that remain are the 
canonical volume ($C=0$, $C_{k}=0$) and surface 
($[Q^{\perp\perp}]=8 \pi G T^{\perp\perp}$, $[Q_{K}^{\perp}]=8 \pi G T^{\perp}_{K}$) 
constraints. At the shell, we are left also
with two equations of motion for the variables $\lambda$ and 
$E(N_{-}^{\perp}/N_{-})$ and the singular constraints $\gamma$-C, 
$\alpha$-C, $\lambda$-C have disappeared. These equations of motion are 
equivalent to the corresponding initial ones, of course via the substitutions
with the new variables. The differentiability
of the final hamiltonian on the final phase space as well as 
the differentiability of the final (canonical) constraints have been proved.\\
These results are important first steps for the quantization of the system 
described in ~\cite{HK}. In order to proceed towards the quantum theory of the
model, the Poisson Algebra of the final constraints remains to be shown equivalent to 
the Dirac Algebra.\\
This problem will be treated in a future work.      
%
%
%
%
%

%
%
%
%
%
\large
\textbf{Acknowledgements} \\ \\
\normalsize
I wish to thank my advisor P.H\'{a}j\'{\i}\v{c}ek for his trust 
and patient. I profited a lot from his great knowledge of General Relativity
and Mathematics. 
I expecially enjoyed the frank but friendly atmosphere of the discussions.\\
I thank my parents for having encouraged me at every stage of this long "trip"
(the PhD).\\
Finally, I thank all the friends that I have met here in Bern during these 
years for the very good and useful time spent together.   
%
%
\end{document}